\documentclass[journal, twocolumn]{IEEEtran}

\usepackage{cite}
\usepackage{amsmath,amssymb,amsfonts}
\usepackage{algorithmic}
\usepackage{graphicx}
\usepackage{textcomp}
\usepackage{subfigure}
\usepackage{booktabs}
\usepackage{multirow}
\usepackage{makecell}
\def\BibTeX{{\rm B\kern-.05em{\sc i\kern-.025em b}\kern-.08em
    T\kern-.1667em\lower.7ex\hbox{E}\kern-.125emX}}
\usepackage{cuted}
\usepackage{color}
\begin{document}

\title{A Radiation Viewpoint of Reconfigurable Reflectarray Elements: Performance Limit, Evaluation Criterion and Design Process}
\author{Changhao Liu, \IEEEmembership{Student Member, IEEE}, You Wu, Songlin Zhou, Fan Yang, \IEEEmembership{Fellow, IEEE}, Yongli Ren, Shenheng Xu, \IEEEmembership{Member, IEEE} and Maokun Li, \IEEEmembership{Senior Member, IEEE}
	\thanks{This work was supported in part by the National Key Research and Development Program of China under Grant No. 2017YFA0700201, in part by ZTE Industry-Academia-Research Cooperation Funds, and in part by THE XPLORER PRIZE. \textit{(Corresponding author: Fan Yang.)}}
	\thanks{The authors are with the Beijing National Research Center for Information Science and Technology (BNRist), Department of Electronic Engineering, Tsinghua University, Beijing 100084, China (e-mail: fan\_yang@tsinghua.edu.cn).}
}

\maketitle

\setcounter{page}{1}

\begin{abstract}
	Reconfigurable reflectarray antennas (RRAs) have rapidly developed with various prototypes proposed in recent literatures. However, designing wideband, multiband, or high-frequency RRAs faces great challenges, especially the lengthy simulation time due to the lack of systematic design guidance. The current scattering viewpoint of the RRA element, which couples antenna structures and switches during the design process, fails to address these issues. Here, we propose a novel radiation viewpoint to model, evaluate, and design RRA elements. Using this viewpoint, the design goal is to match the element impedance to a characteristic impedance pre-calculated by switch parameters, allowing various impedance matching techniques developed in classical antennas to be applied in RRA element design. Furthermore, the theoretical performance limit can be pre-determined at given switch parameters before designing specific structures, and the constant loss curve is suggested as an intuitive tool to evaluate element performance in the Smith chart. The proposed method is validated by a practical 1-bit RRA element with degraded switch parameters. Then, a 1-bit RRA element with wideband performance is successfully designed using the proposed design process. The proposed method provides a novel perspective of RRA elements, and offers a systematic and effective guidance for designing wideband, multiband, and high-frequency RRAs.
\end{abstract}

\begin{IEEEkeywords}
	Metasurfaces, microwave network theory, one-bit, reconfigurable intelligent surfaces, reconfigurable reflectarray antennas.
\end{IEEEkeywords}

\setcounter{figure}{0}
\renewcommand*{\thefigure}{\arabic{figure}}
\renewcommand*{\thetable}{I}

\section{Introduction}
\IEEEPARstart{R}{eflectarrays} are a type of large-aperture antennas that have gained increasing attention in recent years due to their low cost, low profile, light weight, and high performance \cite{reflectarray}. Reconfigurable reflectarray antennas (RRAs) are a specific type of reflectarray that have tunable components in their elements, allowing for dynamic beam steering. Compared to conventional phased arrays, RRAs have numerous benefits such as low power dissipation, high-frequency operation, and large-scale deployment, and are being researched extensively for use in wireless communications, radar detection, remote sensing scenarios, and more \cite{RRA1, RRA2, RRA3}.

Analog RRAs can introduce continuous phase tuning range by loading tunable devices on each element, such as varactor diodes, liquid crystals, ferroelectric materials and phase change materials. To reduce the cost and controlling complexity, digital RRAs are further investigated with discrete reflection phase states. The simplest digital RRA architecture is the 1-bit RRA element, which only has two states and requires only one RF switch, such as PIN diodes and micro-electro-mechanical-systems (MEMSs), to generate a 1-bit phase change. Although 1-bit RRAs have slightly weaker performance than analog RRAs \cite{bit1,bit2}, they are still suitable for most scenarios due to their low cost and simplicity. Therefore, 1-bit RRAs are the mainstream RRA architecture, and numerous 1-bit RRA designs have been presented in recent years.

Researchers have investigated 1-bit RRAs with a single switch on each element at various frequency bands \cite{rra1, rra2, rra3, rra4, rra5, rra8, rra9, rra10,rra11}. In 2012, literature \cite{rra2} designed and fabricated X-band 1-bit RRA element and array, which could achieve dynamic beam control with three directions (-5$^\circ$, 0$^\circ$, 5$^\circ$). In 2016, H. Yang~\emph{et al.} conducted a comprehensive investigation on a Ku-band 1-bit RRA, and a 10$\times$10 RRA based on PIN diodes was fabricated with $\pm50^\circ$ scanning beams \cite{rra3}. In \cite{rra9}, a 26.5-GHz RRA based on MEMS was proposed to switch the beam directions. Additionally, 1-bit RRAs can support beam scanning in millimeter wave band, as demonstrated by a 60-GHz RRA prototype which had a gain of 42 dBi \cite{rra1}, and a W-band 96-GHz RRA based on integrated PIN diodes \cite{rra10}. Recently, a THz-band RRA using high electron mobility transistor (HEMT) switch was designed at 210 GHz in \cite{rra11}. Apart from increasing the operating frequency, there has been a research focus on wideband and multiband RRAs. Literature \cite{rra5} and \cite{rra8} presented wideband RRAs with 1-dB bandwidth over 8.4\% and 22.3\% respectively. Besides, based on different resonant modes, a 1-bit RRA can support dual-frequency operation at X/Ku-band \cite{rra4}.

RRAs are evolving rapidly with various functions, and the research on RRA is moving towards broadband, multiband and high-frequency designs. However, current design methodologies face great challenges when designing these RRAs with high performance. For instance, in millimeter wave and THz band, the switch performances deteriorate, leading to significant reflection loss. Designers are unsure if the current element structure design is optimal at given switch parameters, which calls for a performance limit theory and qualitative evaluation method to assist in judging element performance.
Besides, without clear design targets, designers lack guidance when optimizing RRA structures at desired frequency bands, leading to a lengthy process in trial-and-error simulations. 
For example, in wideband and multiband designs, RRA elements need to operate at multiple resonant modes in desired frequency bands, which leads to significant design complexity. As a result, there is an urgent demand for a simple and efficient design process for RRA elements.

This article presents a systematic design and evaluation method of RRA elements based on the novel radiation viewpoint, extending the findings in our previous conference paper \cite{conf}. In Section II, the single-switch RRA element is viewed as a passive radiation antenna based on two-port network equivalence. The design target, performance limit, and constant loss curves are calculated at given switch parameters. Then, a novel RRA design process based on the proposed method is presented. In Section III, the proposed method is validated by a practical 1-bit design example at THz band. The proposed design method is then applied to design a wideband 1-bit RRA element at C band in Section IV. Finally, the conclusion is drawn in Section V.

\section{Theory of RRA Element from Radiation Viewpoint}

\subsection{Modeling RRA Element Using Two-Port Network Equivalence}

\begin{figure} 
	\centering 
	\subfigure[] { \label{RRAele} 
		\includegraphics[width=0.6\columnwidth]{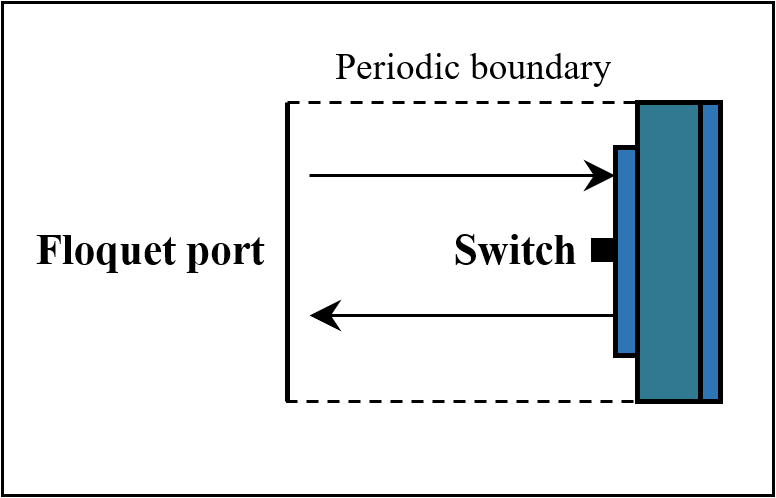} 
	} 
	\subfigure[] { \label{RRAnet} 
		\includegraphics[width=0.6\columnwidth]{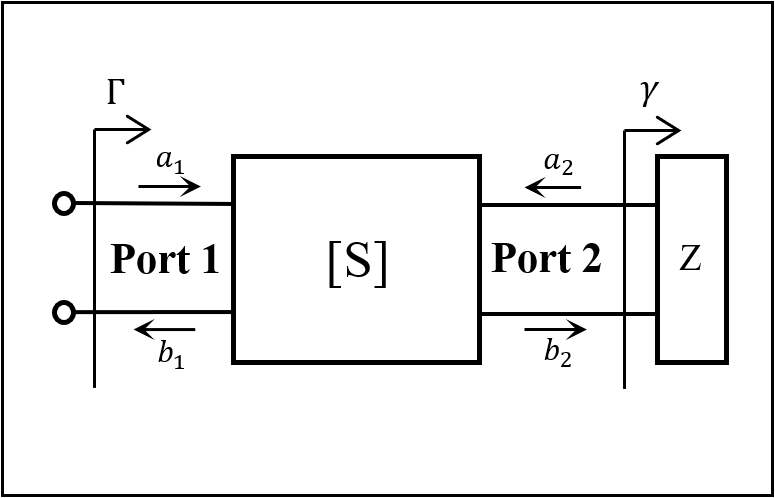} 
	} 
	\caption{The model of an RRA element with a single switch. (a) A schematic diagram of an RRA element. (b) The equivalent two-port network model of the RRA element.} 
	\label{RRAsche} 
\end{figure}

A schematic diagram of an RRA element with a single switch is shown in Fig. \ref{RRAele}. The incident wave is from Floquet port, and reflects backward after encountering the antenna structure with the lumped switch. It is assumed that the majority of energy is reflected with no cross polarization. It is worth noting that there is a reflective ground plane, meaning that no energy is transmitted, and the switch and substrate absorb minor energy. The element is surrounded by a periodic boundary. The different states of the switch can generate reflection coefficients with varying amplitude and phase.

Assuming that the size of the lumped switch is significantly smaller than the operating wavelength $\lambda$, a lumped port can be defined at the switch location. Consequently, the passive element structure can be modeled as a two-port $[S]$ parameter. Port 1 connected to the free space, and port 2 is connected to the switch. The loaded switch is modeled as a variable impedance $Z$, as shown in Fig. \ref{RRAnet}. 

Based on the definition of the scattering matrix \cite{network}, outward wave $[b]$ and inward wave $[a]$ are linked by $[S]$ as follows
\begin{equation}
	\left[\begin{array}{l}
		b_{1} \\
		b_{2} 
	\end{array}\right]
	=
	\left[\begin{array}{ll}
		S_{11} & S_{12} \\
		S_{21} & S_{22}
	\end{array}\right]
	\cdot
	\left[\begin{array}{l}
		a_{1} \\
		a_{2} 
	\end{array}\right].
	\label{bsa}
\end{equation}
The reflection coefficient of the bare switch is expressed as
\begin{equation}
	\gamma = \frac{Z-\eta_0}{Z+\eta_0} = \frac{a_2}{b_2},
\end{equation}
where $\eta_0$ represents the reference characteristic impedance. If the switch has $N$ variable impedance states, then $\gamma$ also has $N$ values, denoted as $\gamma_i$, $i\in {1, ... , N}$. These values are solely determined by the switch parameters.

By utilizing the two-port network to model the RRA element with a single switch, the active element is decoupled into passive structures and the active switch. $[S]$ is used to model general passive antenna structures, while $Z$ is used to model the active lumped switch. This modeling approach can be further extended to all RRA elements. Under the two-port equivalence, the antenna structure can be interpreted as an impedance transforming network, which transforms the impedance of the switch at port 2 to the impedance of the entire element at port 1. The primary design objective for the RRA element is to design a proper impedance transformer that can generate the desired $[S]$ matrix. Subsequently, the desired reflection coefficients from port 1 can be obtained.

\subsection{Theoretical Analysis and the Radiation Viewpoint of RRA Element}

In this part, theoretical derivations are presented to demonstrate that the reflection coefficient $\Gamma$ is related with the reflection coefficient from port 2 ($S_{22}$) and switch parameter ($\gamma$) under the assumption of a lossless passive antenna structure.

First, inserting $a_2 = \gamma b_2$ into (\ref{bsa}) and solving the equation, we obtain the expression
\begin{equation}
	b_1 = S_{11}a_1 + \frac{S_{12}S_{21}\gamma}{1-S_{22}\gamma}a_1.
\end{equation}
Suppose the passive structure is magnet-free and reciprocal, so $S_{21}=S_{12}$. The reflection coefficient from port 1 is
\begin{equation}
	\Gamma =\frac{b_1}{a_1} = S_{11} + \frac{S_{12}^2\gamma}{1-S_{22}\gamma}.
	\label{eq4}
\end{equation}

Here, suppose the passive structure is lossless, and thus the $[S]$ matrix is a unitary matrix:
\begin{equation}
	[S][S]^{H}=I,
\end{equation}
where $[S]^{H}$ is the conjugate transpose matrix of $[S]$. The expression for $[S]$ can be rewritten as
\begin{equation}
	[S]=
	\left[\begin{array}{ll}
		|S_{11}|e^{j\theta_{11}} & |S_{12}|e^{j\theta_{12}} \\
		|S_{21}|e^{j\theta_{21}} & |S_{22}|e^{j\theta_{22}}
	\end{array}\right],
\end{equation}
and under the unitary condition, we can obtain three equations:
\begin{equation}
	\left \{
	\begin{array}{l}
		|S_{11}|=|S_{22}| \\
		|S_{12}|=\sqrt{1-|S_{22}|^2} \\
		2\theta_{12} = \theta_{11}+\theta_{22}+(2k+1)\pi
	\end{array}
	\right.
	\label{eq7}
\end{equation}

Inserting the equations in (\ref{eq7}) into (\ref{eq4}), the reflection coefficient is derived as follows
\begin{equation}
	\begin{aligned}
		\Gamma &= |S_{22}|e^{j\theta_{11}}+\frac{(1-|S_{22}|^2)e^{j(\theta_{11}+\theta_{22}+\pi)}\gamma}{1-|S_{22}|e^{j\theta_{22}}\gamma} \\
		& = e^{j\theta_{11}}\left(|S_{22}|-\frac{(1-|S_{22}|^2)e^{j\theta_{22}}\gamma}{1-|S_{22}|e^{j\theta_{22}}\gamma}\right).
	\end{aligned}
	\label{eq8}
\end{equation}
It is known that RRA element performance is only determined by the relative reflection phase difference and amplitudes. Although (\ref{eq8}) includes $e^{j\theta_{11}}$, this term only affects the absolute reflection phase shift. Therefore, $e^{j\theta_{11}}$ can be ignored and (\ref{eq8}) is then simplified as
\begin{equation}
	\Gamma = \frac{|S_{22}|-e^{j\theta_{22}}\gamma}{1-|S_{22}|e^{j\theta_{22}}\gamma}.
	\label{eq9}
\end{equation}
In (\ref{eq9}), it is clear that the reflection coefficients $\Gamma_i$ at Floquet port 1 are determined by the switch parameters $\gamma_i$ and the structure parameter $S_{22}$. In other words, only the $S_{22}$ parameter in the $[S]$ matrix can affect the performance of an RRA element at a given set of switch parameters. 

\begin{figure} 
	\centering 
	\subfigure[] { \label{understandsca} 
		\includegraphics[width=0.45\columnwidth]{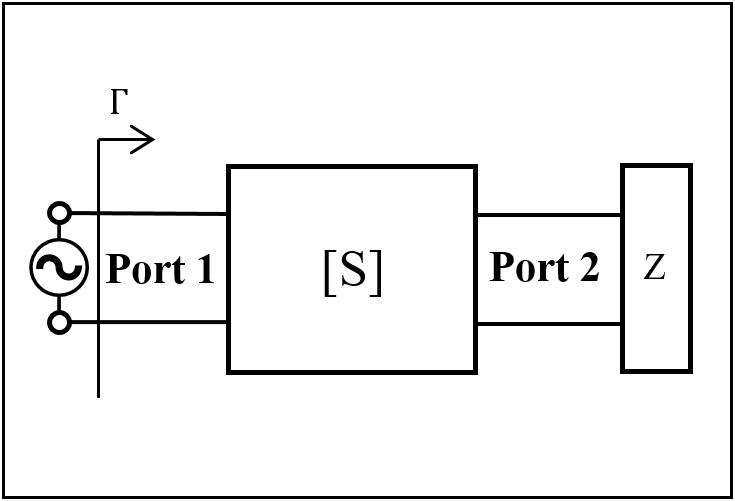} 
	} 
	\subfigure[] { \label{understandrad} 
		\includegraphics[width=0.45\columnwidth]{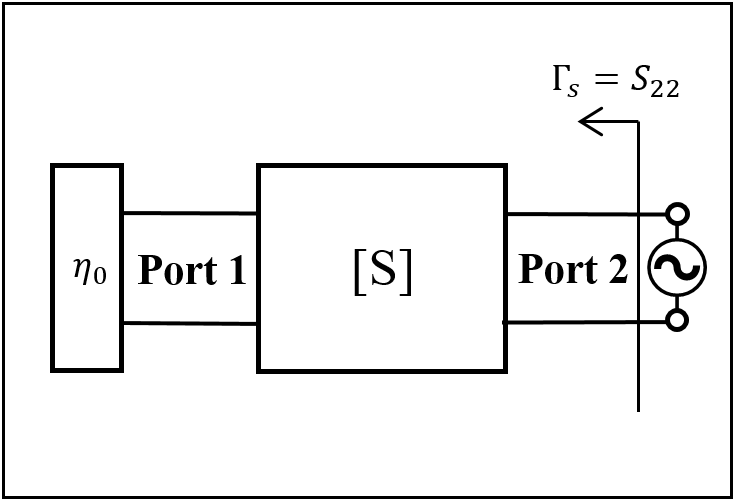} 
	} 
	\caption{Two understandings of an RRA element under network equivalence. (a) Scattering viewpoint. (b) Radiation viewpoint.} 
	\label{understand} 
\end{figure}

Based on the observations above, we compare two ways to understanding an RRA element, as illustrated in Fig. \ref{understand}. The traditional and straightforward approach views the antenna structure with the multi-state switch as a scattering structure. The passive structure and active switch are treated as a single entity that scatters the incident wave, as illustrated in Fig. \ref{understandsca}. In the conventional design process, different $\Gamma_i$ values are simulated and obtained under different switch states $\gamma_i$.

The second way of understanding is the radiation viewpoint, as shown in Fig. \ref{understandrad}. Note that impedance of port 1 is defined as $\eta_0$, which is always loaded with the characteristic impedance in free space, and thus $a_1=0$. Hence, the reflection coefficient from port 2 ($\Gamma_s$) is simply given by
\begin{equation}
	\Gamma_s=\frac{b_2}{a_2}=S_{22}.
\end{equation}
This implies that the structure parameter $S_{22}$ is equal to the reflection coefficient $\Gamma_s$ from the switch port. Therefore, if a lumped port is applied at port 2 as an excitation source and the radiation boundary is connected to port 1, the passive antenna structure only needs to be simulated once to get the response $S_{22}$. Subsequently, the multi-state reflection coefficients $\Gamma_i$ can be directly calculated using (\ref{eq9}). 

The key insight is that Fig. \ref{understandrad} also represents the equivalent network model of a classical single-port radiation antenna. Therefore, under radiation viewpoint, the scattering RRA structure is converted to a classical radiation antenna structure through network equivalence. This allows the systematic and advanced impedance matching techniques used in radiation antenna design to be applied to RRA element design now.

It is noted that the findings presented in this study are based on the assumption of a lossless passive antenna structure. In practice, the substrate is typically lossy, so $[S]$ is not strictly unitary, and $\Gamma$ is related with three parameters in $[S]$ matrix, as derived in (\ref{eq4}). This implies that the radiation viewpoint may not be directly applicable in such cases. But fortunately, in most situations, the loss in the passive antenna structure is usually less than the loss in switches. As a result, the effect of substrate loss on the reflection performance of the antenna is usually minimal. Therefore, designers can initially design the antenna structure using the radiation viewpoint, assuming a lossless antenna, and then verify the reflection performance using the conventional scattering viewpoint, taking into account the antenna loss.

\subsection{The Design Target and Theoretical Performance Limit of RRA Element}
\label{secIIC}

An advantage of the proposed radiation viewpoint of the RRA element is to set a clear design target with a corresponding performance limit. To determine the design target and the theoretical limit of RRA element performance at given switch parameters, an equivalent reflection amplitude (ERA) is introduced in this part.

The reflection coefficient from port 1 can be expressed as
\begin{equation}
	\Gamma_i = A_ie^{j\varphi_i} \quad i\in {1, ... , N},
\end{equation}
where $A$ is the reflection amplitude and $\varphi$ is the reflection phase. Based on the set of $\Gamma_i$ values, ERA is defined as
\begin{equation}
	\text{ERA}=\frac{1}{2\pi}\int_{0}^{2\pi}\max\limits_{i}\{A_i\cos(\phi-\varphi_i)\} \mathrm{d} \phi.
	\label{eq12}
\end{equation}
Since $\Gamma$ is a function of $S_{22}$ at given switch parameters according to (\ref{eq9}), ERA is also a function of $S_{22}$. ERA considers both reflection amplitude and phase difference to comprehensively describe the reflection performance of an RRA element. For instance, in the ideal case, the reflection amplitudes under all cases are 1, and the reflection phase can be continuously chosen, resulting in $\text{ERA}=1$, indicating no equivalent reflection loss for the ideal element. Actually, the expression of ERA is defined to statistically maximize the gain of the reflectarray, and derivation details are provided in Appendix \ref{appdA}.

At a given set of switch parameters, different antenna structures result in different $S_{22}$ values and thus generate different ERAs. The theoretical performance limit (PL) of an RRA element is defined as the maximum possible value of ERA. Mathematically, it is expressed as
\begin{equation}
	\text{PL} = \underset{S_{22}}{\operatorname{max}}\{\text{ERA}(S_{22})\}.
	\label{pl}
\end{equation}
On the other hand, $S_{22}^t$ is the optimal solution of $S_{22}$ that maximizes the ERA and can be used as a design target:
\begin{equation}
	S_{22}^t = \underset{S_{22}}{\operatorname{argmax}}\{\text{ERA}(S_{22})\}.
	\label{s22t}
\end{equation}

In order to attain specific PL and $S_{22}^t$ values based on the provided switch parameters, a straightforward numerical enumeration approach can be employed. This approach enables a comprehensive coverage of the entire Smith chart by roughly sampling $S_{22}$. Considering that $S_{22}$ possesses solely two degrees of freedom, namely $|S_{22}|\in[0, 1]$ and $\theta_{22}\in[0, 2\pi]$, the numerical traversal calculation to find the optimized $S_{22}^t$ can be completed in less than a second using the Matlab.

\begin{figure}[!t]
	\centerline{\includegraphics[width=0.45\columnwidth]{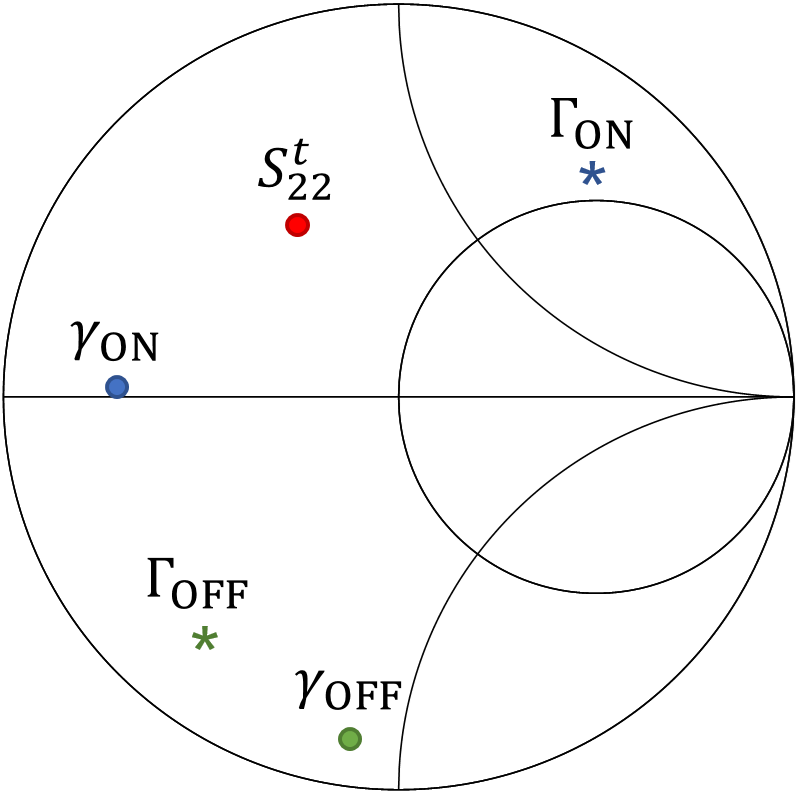}}
	\caption{A conceptual diagram of switch parameters ($\gamma_\text{ON}, \gamma_\text{OFF}$), reflection coefficients ($\Gamma_\text{ON}, \Gamma_\text{OFF}$) and design target ($S_{22}^t$) in the Smith chart. The switch parameters are transformed by the optimal structure parameter into the optimal reflection coefficients.}
	\label{Z1bit}
\end{figure}

Particularly, for 1-bit RRA elements, $\gamma$ has two states: $\gamma_\text{ON}$ and $\gamma_\text{OFF}$. The optimal solution $S_{22}^t$ can be obtained and plotted in the Smith chart, as illustrated in Fig. \ref{Z1bit}. In this example, the original switch parameters $\gamma_\text{ON}, \gamma_\text{OFF}$, which do not have 1-bit phase difference, are transformed by $S_{22}^t$ into two reflection states $\Gamma_\text{ON}$ and $\Gamma_\text{OFF}$ that have 1-bit phase difference and result in the maximum ERA.

In Fig. \ref{Z1bit}, it is also observed that the optimal reflection coefficients of 1-bit elements are always in opposite positions, namely, $\Gamma_\text{ON}=-\Gamma_\text{OFF}$. This observation agrees with the previous design experience, which suggests that when the reflection phase difference is near 180$^\circ$ and the reflection amplitudes are close, the 1-bit element performance is usually optimal. Furthermore, based on this simplified condition that $\Gamma_\text{ON}=-\Gamma_\text{OFF}$, an analytical solution for $S_{22}^t$ are obtained in (\ref{eqsolvtheta22}) and (\ref{eqsolvs22}). The derivations are shown in Appendix \ref{appdB}.

In summary, using ERA to evaluate the element performance quantitatively, the PL of RRA element can be pre-calculated using (\ref{pl}) once the switch parameters are given. This can help designers to predict the reflection loss and aperture gain before designing specific antenna structures. In addition, the design target for the passive structure, $S_{22}^t$, is clear and unique according to (\ref{s22t}) or (\ref{eqsolvtheta22}) and (\ref{eqsolvs22}). When optimizing a specific structure, if the actual reflection coefficient from port 2 ($S_{22}$) matches with $S_{22}^t$ at the desired frequency, the designed RRA element can reach the PL. Moreover, since the scattering element can be viewed as a radiation antenna and the goal is to match the structure impedance with $S_{22}^t$, the design method and parameter tuning process of RRA element is the same as the impedance matching techniques used for classical antennas.

\subsection{An Evaluation Criterion at Mismatched Points}
\label{secIID}

In practical RRA design, $S_{22}$ varies with frequency, which may not always match the optimal design target $S_{22}^t$ within a certain frequency band, making it not feasible for the element to exactly meet the PL over the band. Therefore, it is useful to have a quantitative criterion to evaluate the relative loss when $S_{22}$ does not match $S_{22}^t$.

\begin{figure} 
	\centering 
	\subfigure[] { \label{CLC} 
		\includegraphics[width=0.45\columnwidth]{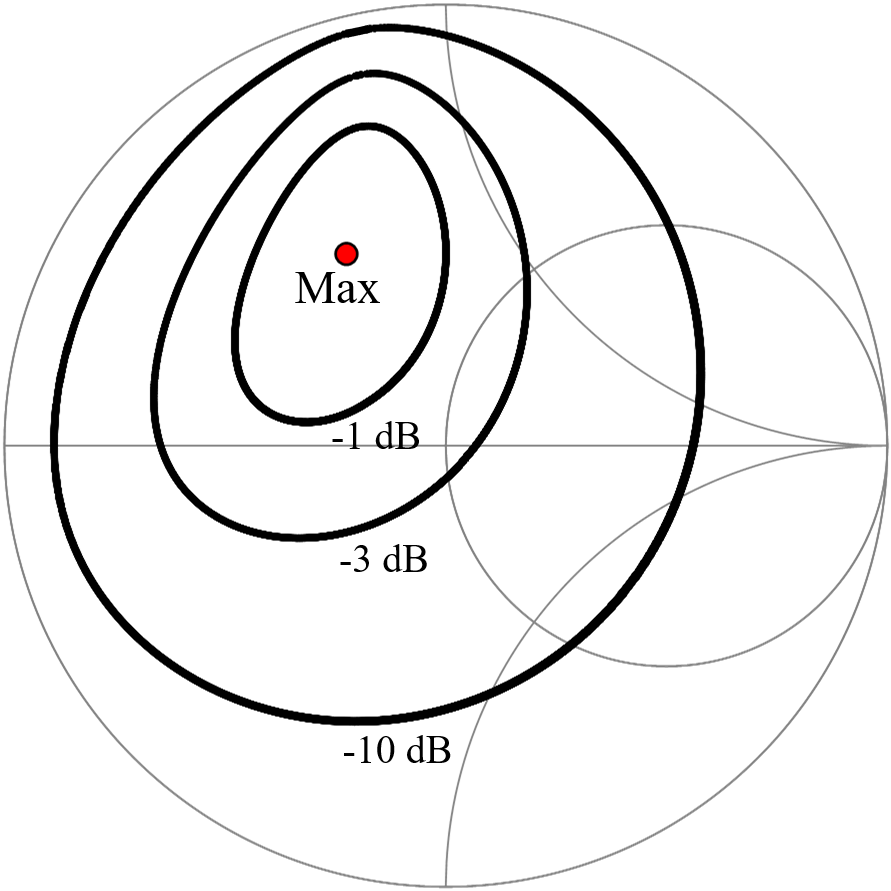} 
	} 
	\subfigure[] { \label{CLCcomp} 
		\includegraphics[width=0.45\columnwidth]{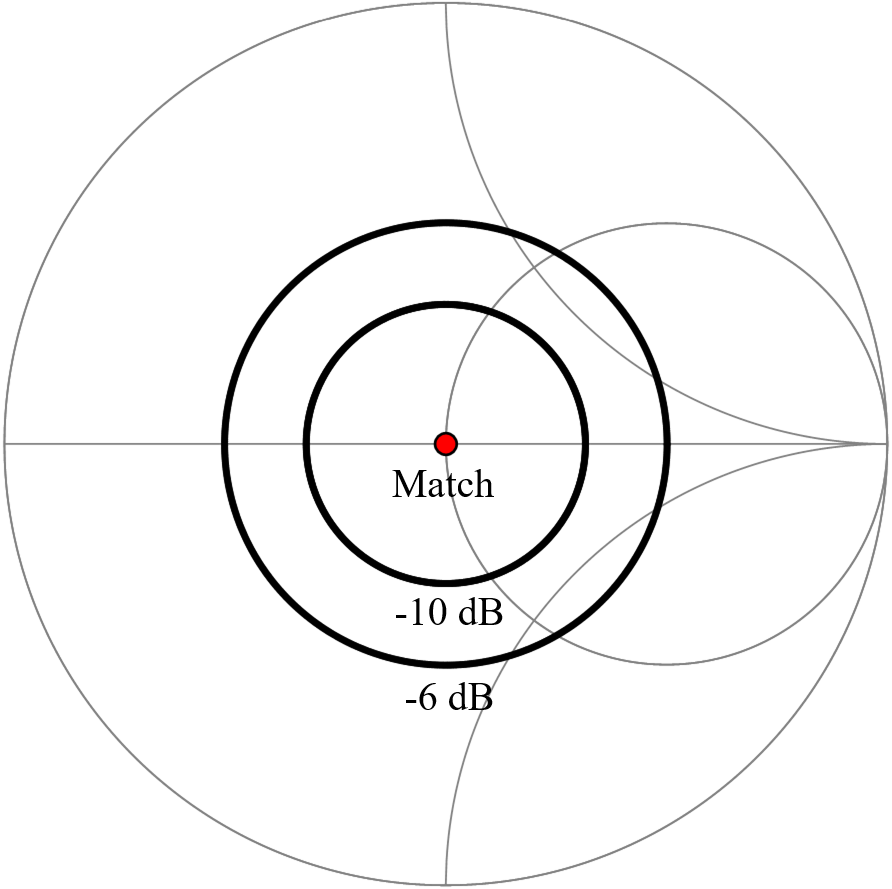} 
	} 
	\caption{Constant curves of RRA elements and classical antennas respectively. (a) A conceptual diagram of the constant loss curves (CLCs) of an RRA element using the radiation viewpoint in the Smith chart, where each certain curve has the same ERA. The relative loss values are compared with the PL value. (b) The constant reflection coefficient curves of a classical radiation antenna in the Smith chart, where each certain curve has the same reflection amplitude.}
	\label{CLCs} 
\end{figure}

Actually, the ERA defined in (\ref{eq12}) can be used to quantitatively evaluate the element performance of the RRA element for all $S_{22}$. This also allows us to calculate the relative reflection loss at mismatched points using (\ref{eq12}). Intuitively, at given switch parameters, every point of $S_{22}$ in the Smith chart corresponds to a certain ERA, and every $S_{22}$ point with the same ERA forms a closed curve in the Smith chart. By selecting different ERA levels, a series of curves called constant loss curves (CLCs) can be generated, as illustrated in Fig. \ref{CLC}. If the $S_{22}$ varies inside a certain CLC over the band, the relative reflection loss of this RRA element is lower than the loss level of the CLC within the band. 
Besides, the outer CLC has a larger constant loss level, which means that more $S_{22}$ values are eligible if the target loss level is larger. Therefore, the design restrictions on the RRA element are looser if the element's performance deterioration can be tolerated. Particularly, as the CLC level decreases, the phase difference becomes narrow or the reflection magnitude loss increases. The introduction of CLCs provides an intuitive aid of designing and evaluating RRA elements over a frequency band.

The proposed quantitative evaluation method at mismatched points extends the proposed method from a single frequency point to a frequency band, which is useful for designing and evaluating wideband or multiband RRA elements in a quantitative manner. Furthermore, the pre-calculated CLCs in the Smith chart can provide an intuitive standard to evaluate the wideband performance of RRA element, similar to -10 dB or -6 dB impedance bandwidth criteria commonly used for performance evaluation in radiation antennas, as shown in Fig. \ref{CLCcomp}. 

It is worth mentioning that the pre-calculated parameters, namely PL, $S_{22}^t$, and CLCs, are only determined by the switch parameters ($\gamma_i$). Moreover, the frequency characteristics of these pre-calculated parameters are solely reliant on the switch parameters as well. Given that the switch parameters typically exhibit stability across a wide operating band range, the pre-calculated parameters are also frequency-invariant in a broad band.

\subsection{The Proposed Design Process}

Using the radiation viewpoint to interpret the RRA element, a novel design process is generalized, which can design and evaluate the RRA elements step by step. 
\begin{enumerate}
	\item \textbf{Pre-calculation}. Pre-calculate the PL, design target $S_{22}^t$ and CLCs based on given switch parameters.
	\item \textbf{Passive antenna structure design}. Design the passive structure, simulate it, and optimize parameters to obtain the reflection coefficient $S_{22}$ from lumped port 2. 
	\item \textbf{Evaluation}. Compare $S_{22}$ with $S_{22}^t$ and certain CLCs in the Smith chart to obtain the central frequency and operation bandwidth.
	\item \textbf{Verification}. Replace the lumped port with lumped RLC series of the multi-state switch and verify the reflection amplitude and phase performance of the entire element. Post-calculate the ERA using the simulation results and (\ref{eq12}), and obtain the final results of element performance.
\end{enumerate}

For comparison, the conventional design process based on scattering viewpoint is also summarized below:
\begin{enumerate}
	\item \textbf{Element design}. Design the structure with the embedded lumped switch, simulate it, and optimize parameters to get the reflection coefficients $\Gamma$ of the entire element from Floquet port 1 under different states respectively.
	\item \textbf{Evaluation}. Compare the reflection phase difference and reflection amplitudes at different states.
\end{enumerate}

Compared to the conventional scattering method, the proposed radiation method requires more preparation before designing a specific element, but it allows designers to estimate the element performance and set a clear target before designing. During the design process, impedance matching techniques of the classical radiation antenna design can be used for designing passive structures and tuning parameters. Both the intuitive Smith chart and the quantitative ERA can be used for evaluation. After finishing the impedance matching process, the switch is loaded onto the optimized structure for verification, and the entire element is simulated using the scattering analysis to get the final reflection coefficients from port 1.

Instead, the conventional method is more straightforward, but it requires the same structure to be simulated for multiple times to get the reflection coefficients under different states, which is time-consuming. More importantly, the parameter optimization lacks clear guidance, which is a blind tuning process.

\subsection{Discussion}

	The differences between the conventional method and the proposed method are summarized in Table \ref{tab:compare}. The conventional method views the RRA element as a scattering structure, which is a natural and straightforward perspective. However, it falls short in determining the upper limit of the element performance. Furthermore, it relies on phase difference and the magnitude loss at multiple states as its design target, resulting in ambiguity and complexity. Consequently, the parameter tuning process lacks clear guidance and heavily relies on the designers' experiential knowledge, making it highly empirical.

Compared with the conventional method, the proposed method uses the radiation viewpoint. By converting the scattering structure to a radiation antenna, it can pre-calculate the performance limit, and set a clear and unique design target before specific designs. Besides, this method uses the ERA and CLCs for evaluation, which is both quantitative and intuitive. As a result, this method provides a systematic framework for parameter tuning in the design of RRA elements. Notably, leveraging the advanced impedance matching techniques developed in classical radiation antennas can greatly aid in the design and optimization of RRA structures.

\renewcommand*{\thetable}{I}
\begin{table}[!t]
	\centering
	\caption{Comparison between the conventional method and the novel method}
	\renewcommand\arraystretch{1.5}
	\setlength\tabcolsep{0.6mm}{
		\resizebox{\linewidth}{!}{
			\begin{tabular}{lcc}
				\hline
				\hline
				Terms & Conventional method & \textbf{Proposed method}      \\ \hline
				Point of view & Scattering structure & Radiation antenna \\				\hline
				Performance limit & Unknown & Known \\				\hline
				Design target & Complex & Clear and unique \\				\hline
				Evaluation criterion & \makecell[c]{Phase difference \\ and magnitude loss} & ERA and CLCs \\				\hline
				Design process & Empirical & \makecell[c]{Systematic \\(impedance matching)} \\
				\hline \hline
			\end{tabular}
		}
	}
	\label{tab:compare}
\end{table}

Specifically, the proposed radiation viewpoint has significant advantages over the conventional scattering viewpoint when designing wideband, multiband, or high-frequency RRA elements. Conventionally, based on the scattering viewpoint, it is difficult to obtain the on-demand reflection performance within specific frequency bands. But with the novel radiation method, impedance matching techniques developed for classical antennas can be applied to design structures with the desired response over a wide range of bands.

Besides, as frequency increases into millimeter wave and THz ranges, the switch performance degrades, making it hard to design structures with optimal reflection coefficients using the conventional scattering viewpoint. Instead, the proposed radiation method can pre-calculate the design target and PL for any switch parameters, making high-frequency RRA design as straightforward as microwave RRA design.

Moreover, the proposed method can also model other types of active metasurface elements using a multi-port microwave network, allowing for investigation beyond single-switch RRA elements. For example, literature \cite{rta} uses the three-port network to model the reconfigurable transmitarray antenna (RTA) element with a single switch, showing that designing a single-switch RTA element is not possible. In the future, various works can be done following the network equivalence path, such as RRA elements under oblique incidence \cite{micromodel}, analog RRA elements using varactors, RRA elements with lossy substrates, amplifying RRA elements using transistors, 2-bit RRA elements using multiple switches, RRA elements related with polarization, RTAs with two switches, active frequency selective surfaces (AFSSs), and more.

\section{Theory Validation: A THz RRA Element with an Imperfect Switch}

In this section, we provide a practical example at the THz band to demonstrate the feasibility of the proposed method. In this band, switch parameters often deteriorate, making it difficult to design RRA elements. Here, we apply the radiation viewpoint to examine the effectiveness of a reported design in the THz band.

\subsection{The Basic Setups} 

\begin{figure}[!t]
	\centering 
	\subfigure[] { 
		\includegraphics[width=0.4\columnwidth]{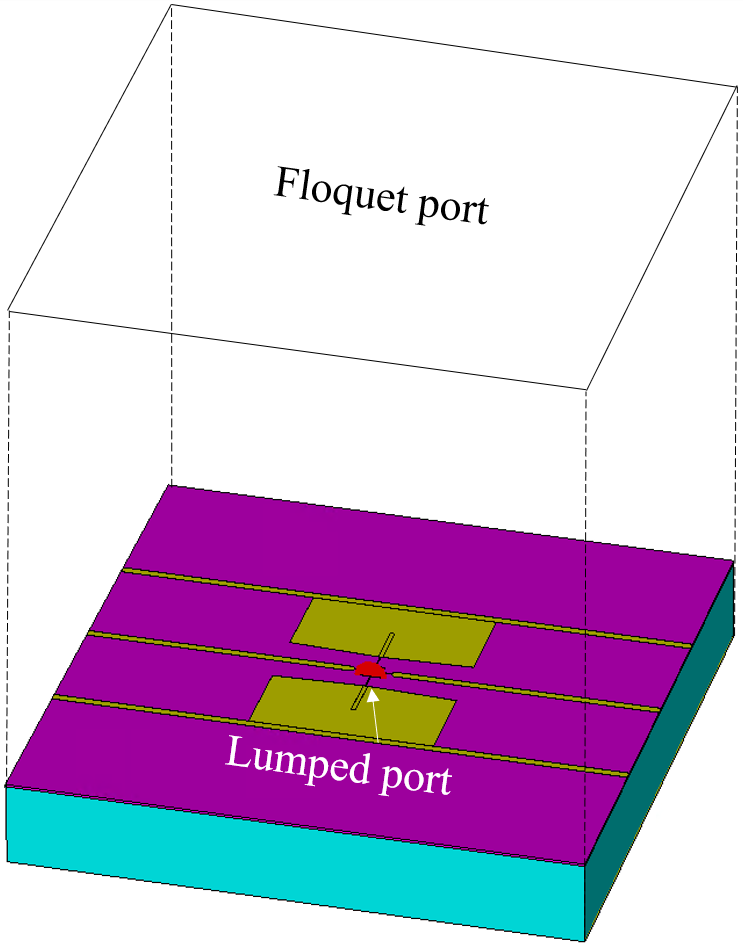} 
	} 
	\subfigure[] { 
		\includegraphics[width=0.7\columnwidth]{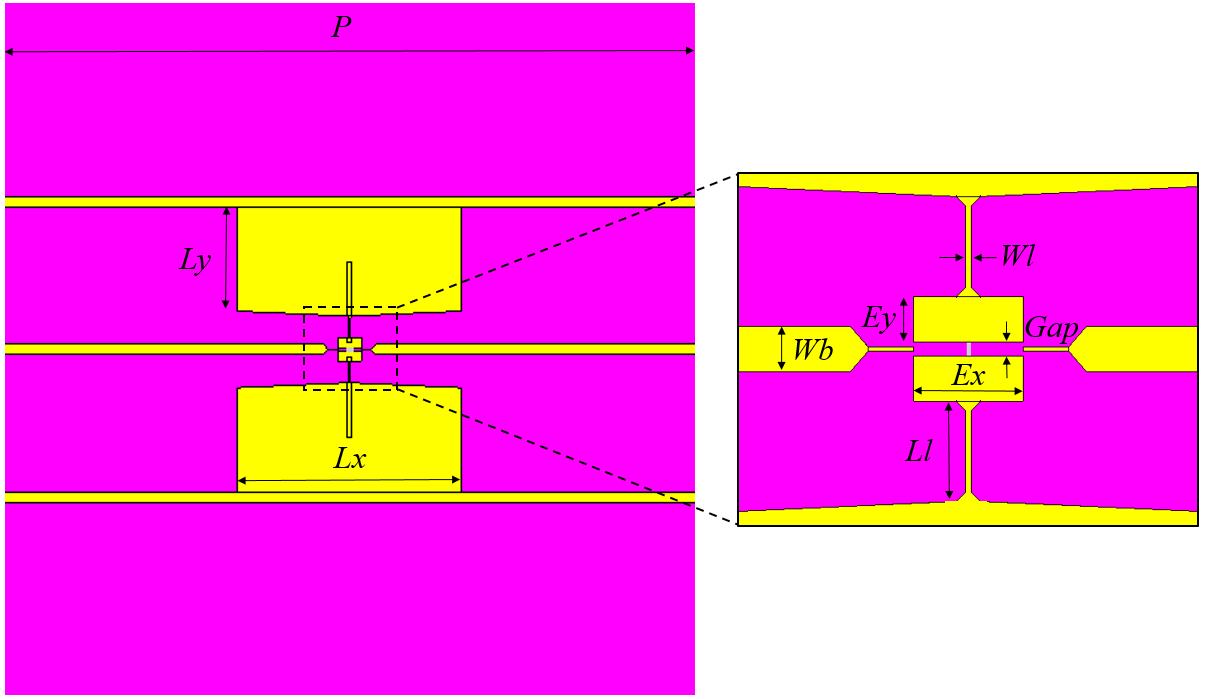} 
	} 
	\caption{The simulated structure at THz band. The yellow area is the metallic pattern, and the pink area is the substrate. (a) The 3D view of the structure.  (b) The top view of the structure. $P=700$ um, $Lx=225$ um, $Ly=105$ um, $Gap=3$ um, $Wb=10$ um, $Ll=18$ um, $Wl=1$ um, $Ex=24$ um, $Ey=10$ um.}
	\label{220Gmodel}
\end{figure}

A 1-bit RRA operating in the THz band is reported in \cite{rra11}. According to the literature, this RRA uses an HEMT switch and can operate up to 200 GHz with a loss of approximately 8 dB in reflection amplitude. Here, these results are used to demonstrate the validity of the proposed method. To do so, a simulation model is built in CST Studio, as shown in Fig. \ref{220Gmodel}, which is the same as the setups described in \cite{rra11}. The substrate is a 100-um-thick sapphire backed by a reflective ground plane. In the radiation method, the lumped switch is replaced by a lumped port at the center of element to excite the patch and obtain the reflection coefficient. The top surface is connected to a Floquet port with a characteristic impedance of 377 $\Omega$, and the element is surrounded by periodic boundaries. 

\subsection{The Calculated Parameters} 

\begin{figure}[!t]
	\centerline{\includegraphics[width=0.5\columnwidth]{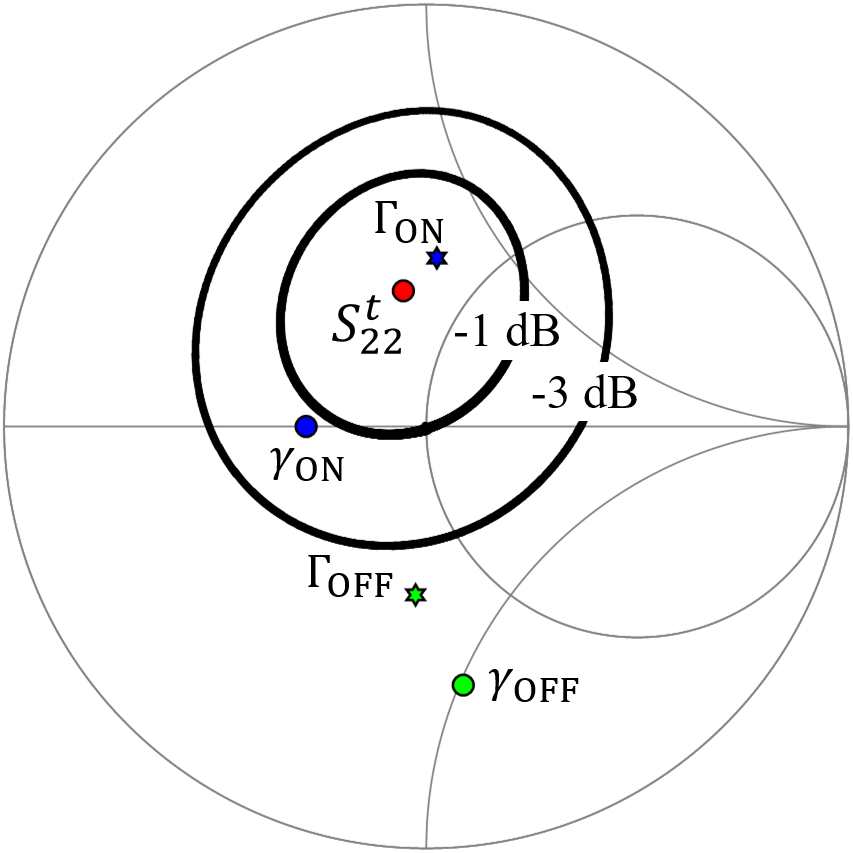}}
	\caption{The parameters pre-calculated from the proposed method.}
	\label{220GS1bit}
\end{figure}

Before simulation, the auxiliary parameters are pre-calculated solely based on the parameters of the switch. The in-house HEMT switch is modeled as lumped RC series at 207 GHz in \cite{rra11}, with $R_{\rm{ON}}=$ 210 $\Omega$; $R_{\rm{OFF}}=$ 192.5 $\Omega$, $C_{\rm{OFF}}=$ 2 fF. At the given switch parameters, pre-calculation is made at given switch parameters with the design target of $S_{22}^t=0.33e^{1.74j}$, as shown in Fig. \ref{220GS1bit}. Unlike high-performance switches in microwave band, the switch parameters deteriorate in THz band, resulting in the calculated PL of only -11.8 dB. As demonstrated in Appendix \ref{appdA}, the 1-bit phase quantization loss is 3.9 dB, resulting in a maximum reflection amplitude is -7.9 dB after removing the phase quantization loss.

\subsection{Element Simulation and Theory Validations} 

\begin{figure}[!t]
	\centerline{\includegraphics[width=0.5\columnwidth]{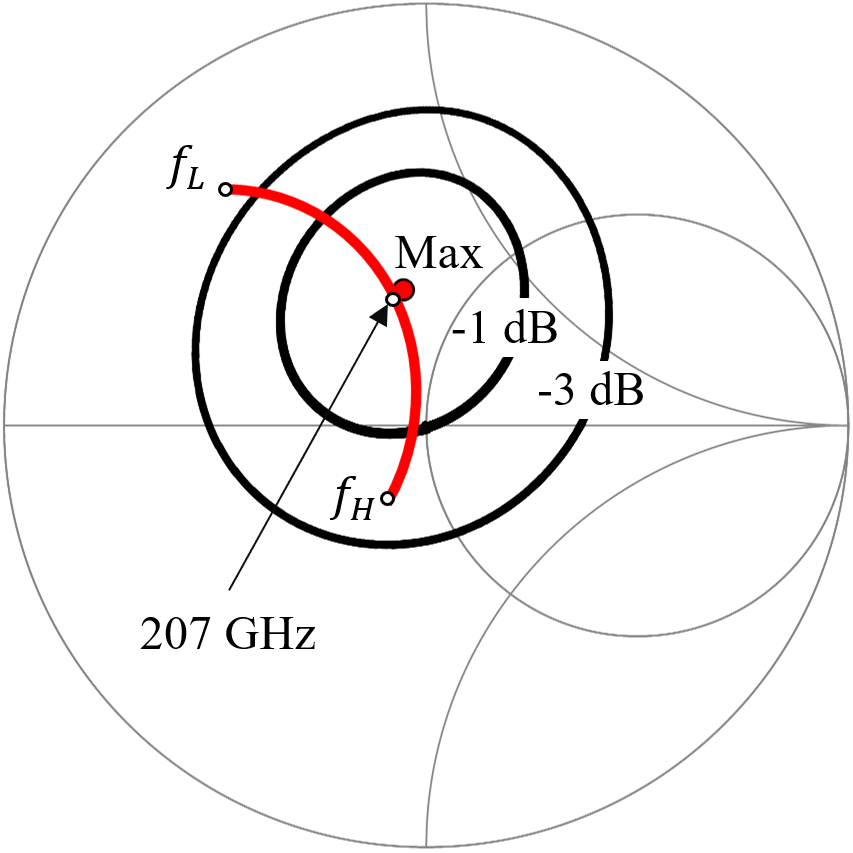}}
	\caption{The simulated $S_{22}(f)$ curve of the structure (red curve) and the pre-calculated parameters, where $f_L=$ 190 GHz and $f_H=$ 220 GHz.}
	\label{220GS22}
\end{figure}

The element is simulated using CST Studio from 190 GHz to 220 GHz. The reflection coefficient $S_{22}(f)$ is obtained from the lumped port and plotted against the pre-calculated parameters in Fig. \ref{220GS22}. It can be observed that the $S_{22}(f)$ curve is in close proximity to the $S_{22}^t$ point at 207 GHz, indicating that the 1-bit RRA element approaches the PL at 207 GHz. Despite the high loss of the HEMT switch, the element is designed with the optimal performance at 207 GHz, which verifies it is a successful design at THz band.

\begin{figure} 
	\centering 
	\subfigure[] { \label{220Gcompareamp} 
		\includegraphics[width=0.6\columnwidth]{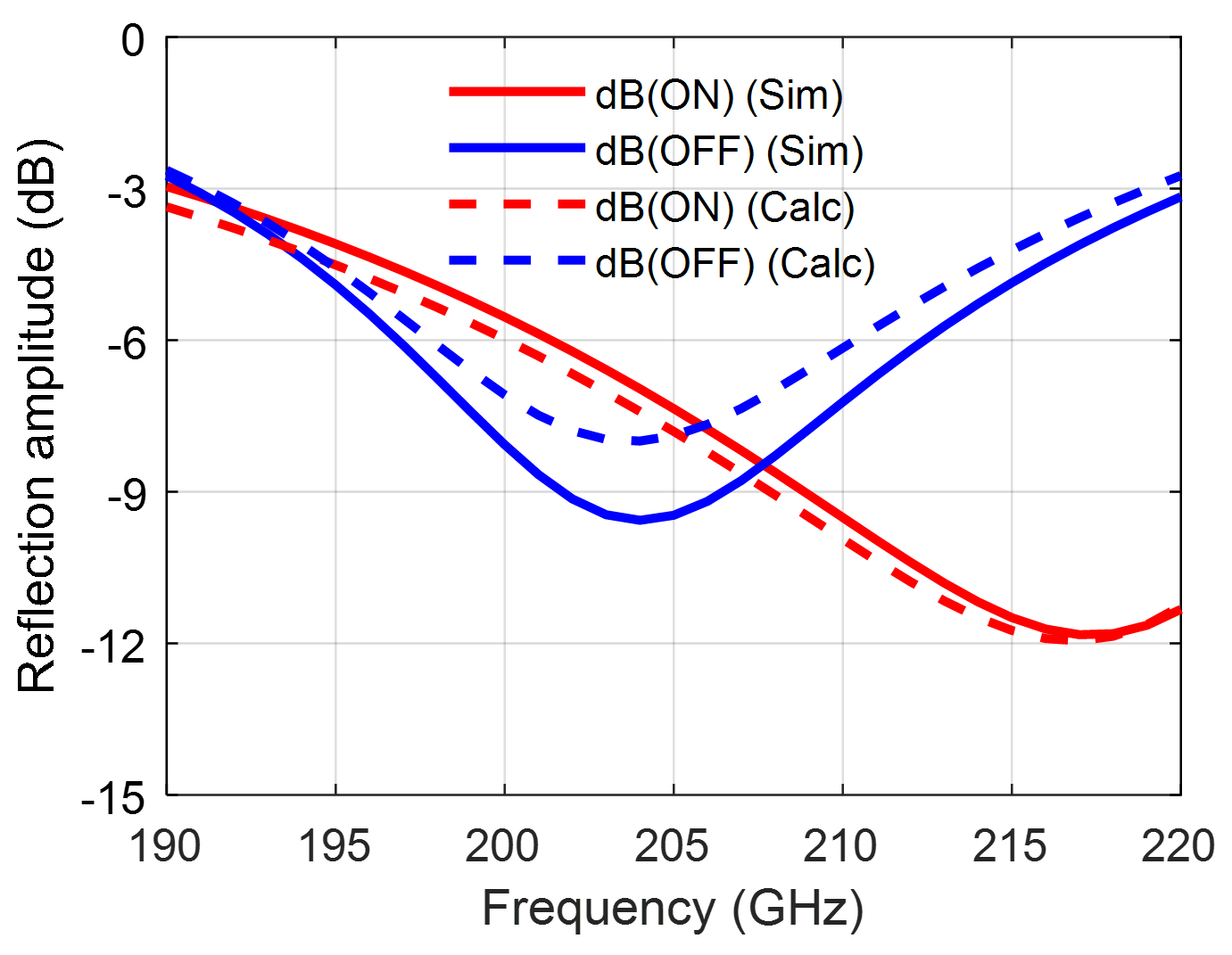} 
	} 
	\subfigure[] { \label{220Gcomparephase} 
		\includegraphics[width=0.6\columnwidth]{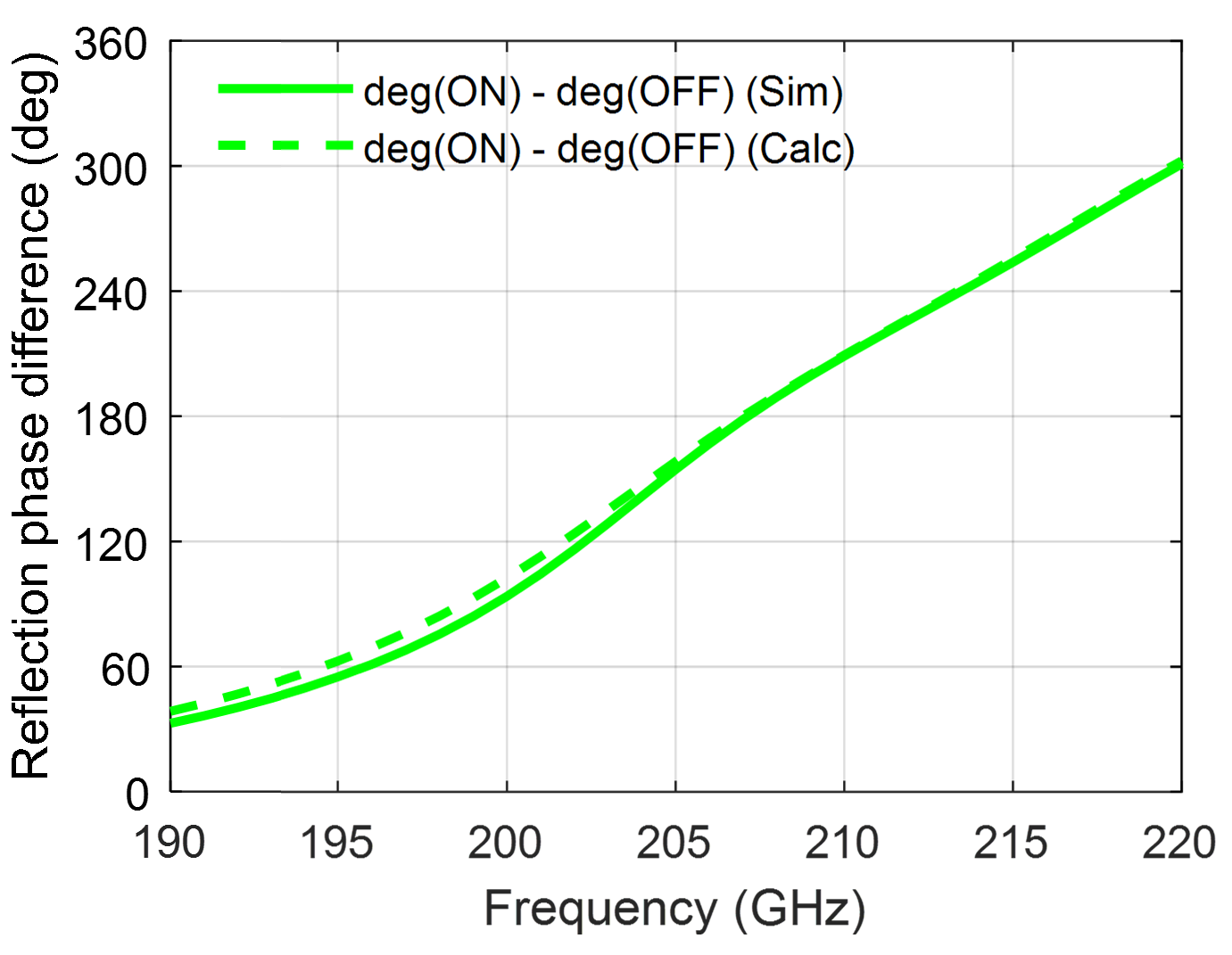} 
	} 
	\caption{The calculated and simulated (a) reflection amplitude and (b) reflection phase difference.} 
	\label{220Gcompare} 
\end{figure}

To verify the proposed method, we compare the calculated results with simulation results obtained using the conventional scattering method. As plotted in Fig. \ref{220Gcompare}, the amplitude and phase difference of both methods match with each other, where the phase difference is 180$^\circ$ at 207 GHz and the reflection amplitudes are about -8.3 dB at 207 GHz. The pre-calculated maximum amplitude limit is -7.9 dB, and the additional loss is due to the substrate loss. The simulated loss is close and within the PL, which demonstrates the validity of the proposed theory, and the PL can help to estimate the element performance before designing the antenna structure.

\section{Theory Application: A Wideband RRA Element}

In this section, we apply the proposed radiation viewpoint to design a wideband 1-bit RRA element at C band, to demonstrate its ability to engineer the frequency-domain features of RRA elements in an effective way.

\subsection{Pre-Calculation}

\begin{figure}[!t]
	\centerline{\includegraphics[width=0.6\columnwidth]{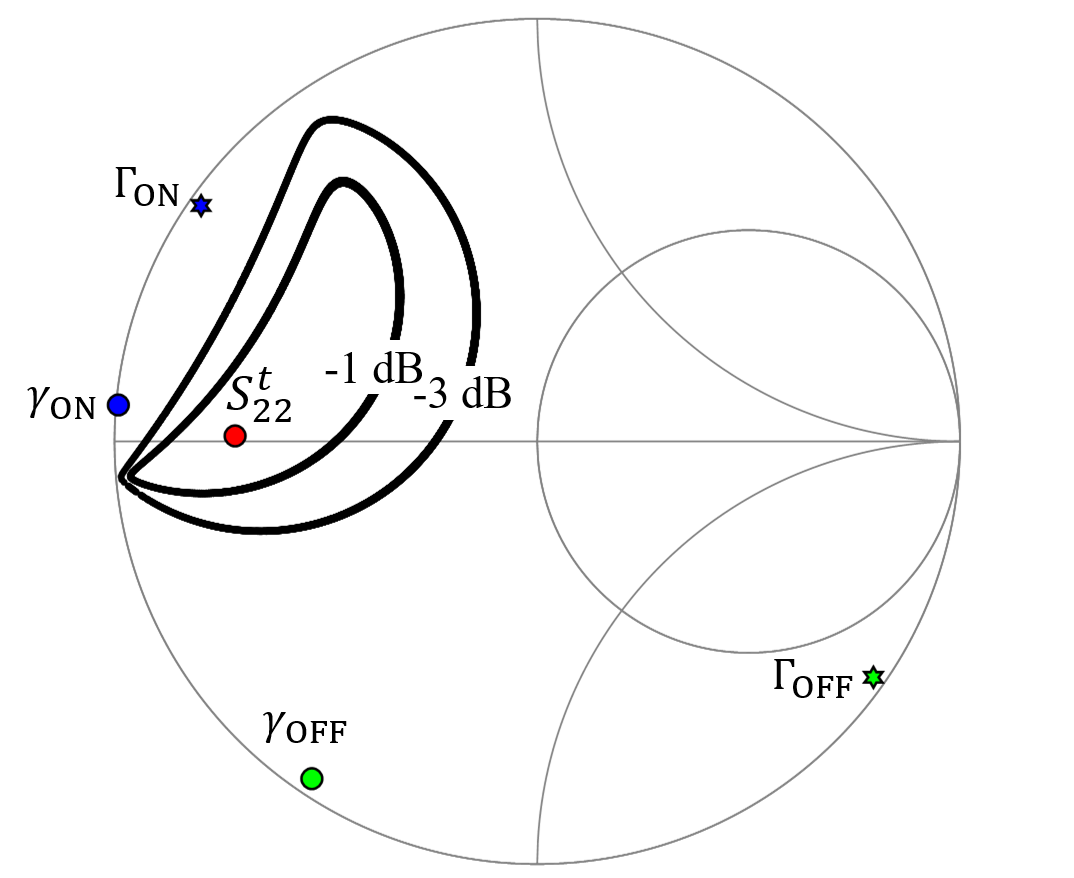}}
	\caption{The pre-calculated parameters based on switch parameters at C band.}
	\label{58Gpara}
\end{figure}

A Skyworks SMP1340 PIN diode is used for this design, which is modeled as lumped RLC series at around 5.8 GHz: $R_{\rm{ON}}=$ 1 $\Omega$, $L_{\rm{ON}}=$ 450 pH; $R_{\rm{OFF}}=$ 10 $\Omega$, $L_{\rm{OFF}}=$ 450 pH, $C_{\rm{OFF}}=$ 126 fF, with reference to \cite{PINpara}. According to the proposed design process, the auxiliary parameters are pre-calculated at these switch parameters at 5.8 GHz. The calculated PL, including the phase quantization loss, is -4.2 dB, so the reflection amplitude loss limit is about 0.3 dB after taking off the 1-bit phase quantization loss of near 3.9 dB. The optimal design target is numerically found as $S_{22}^t=0.71e^{3.1j}$. Additionally, the 1-dB and 3-dB CLCs are also calculated. As an intuitive illustration, the pre-calculated parameters are shown in Fig. \ref{58Gpara}.

\subsection{Structure Design and the Parameter Tuning Process}
\label{Cdesign}
\begin{figure}[!t]
	\centerline{\includegraphics[width=0.5\columnwidth]{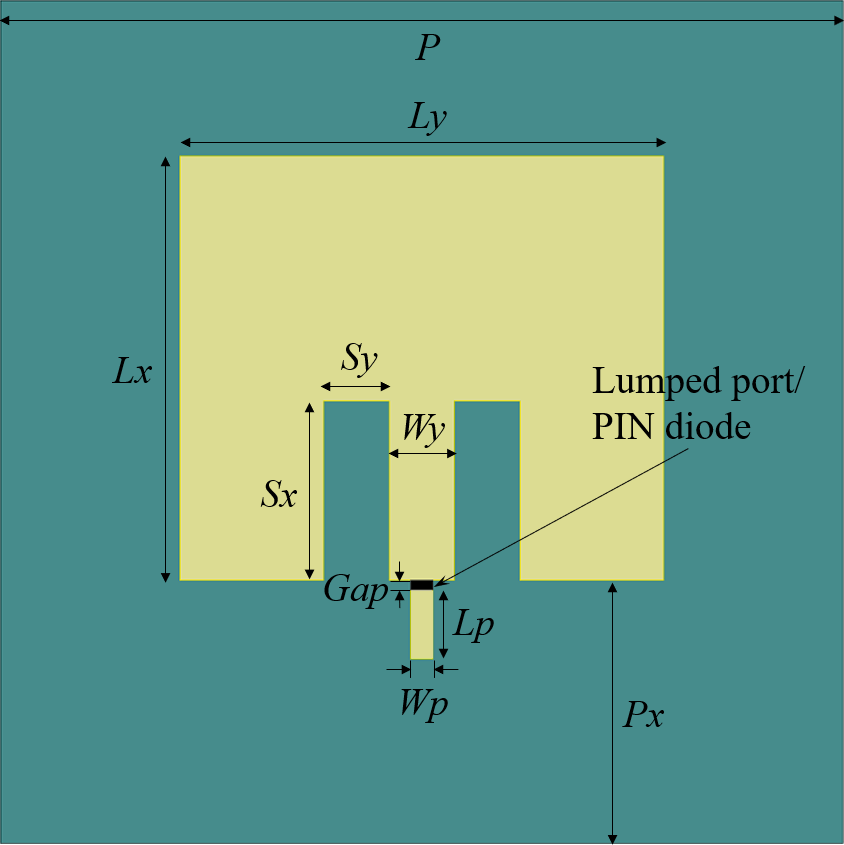}}
	\caption{The wideband E-shaped antenna structure. The light yellow area is the designed metallic pattern, and the dark green area is the substrate. $P=25.8$ mm, $Gap=0.3$ mm, $Wp=0.7$ mm, $Lp=2.1$ mm, $Lx=13$ mm, $Ly=14.8$ mm, $Sx=5.5$ mm, $Sy=2$ mm, $Wy=2$ mm, $Px = 8$ mm.}
	\label{58Gmodel}
\end{figure}

For this wideband RRA design example, a classical E-shaped wideband antenna structure in \cite{Eshape} is used as the passive structure pattern. The designed structure is shown in Fig. \ref{58Gmodel}, with parameters labeled in the figure. The single-layer substrate is Taconic RF-35 ($\epsilon$ = 3.5, tan$\delta$ = 0.0018) with a thickness of 1.52 mm and is backed by a reflective metallic ground plane. The E-shaped patch antenna with the pad is on the front side. Based on the radiation viewpoint, a lumped port with an equal size to the PIN diode is loaded at the gap to excite the patch. The element is surrounded by periodic boundaries. The structure is simulated in Ansys HFSS, from 4 GHz to 7 GHz.

\begin{figure}[!t]
	\centering
	\subfigure[] { 
		\includegraphics[width=0.45\columnwidth]{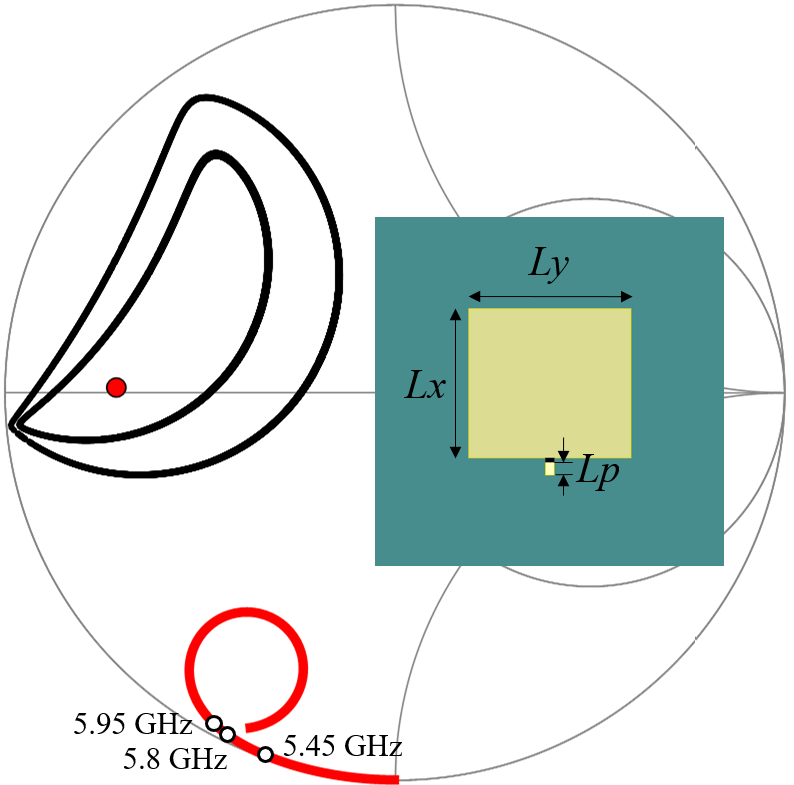} 
	} 
	\subfigure[] { 
		\includegraphics[width=0.45\columnwidth]{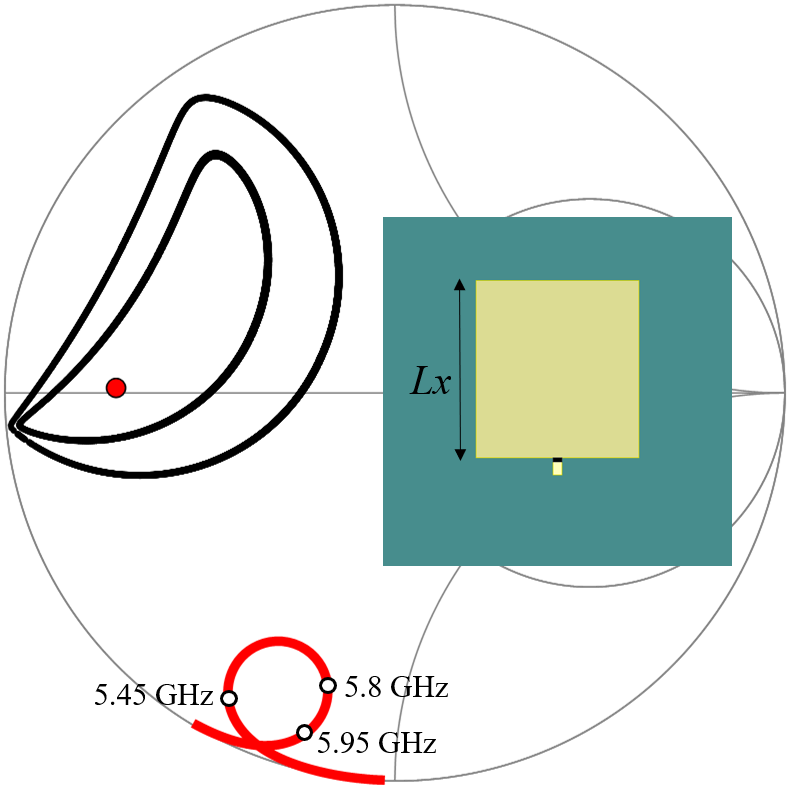} 
	} 
	\subfigure[] { 
		\includegraphics[width=0.45\columnwidth]{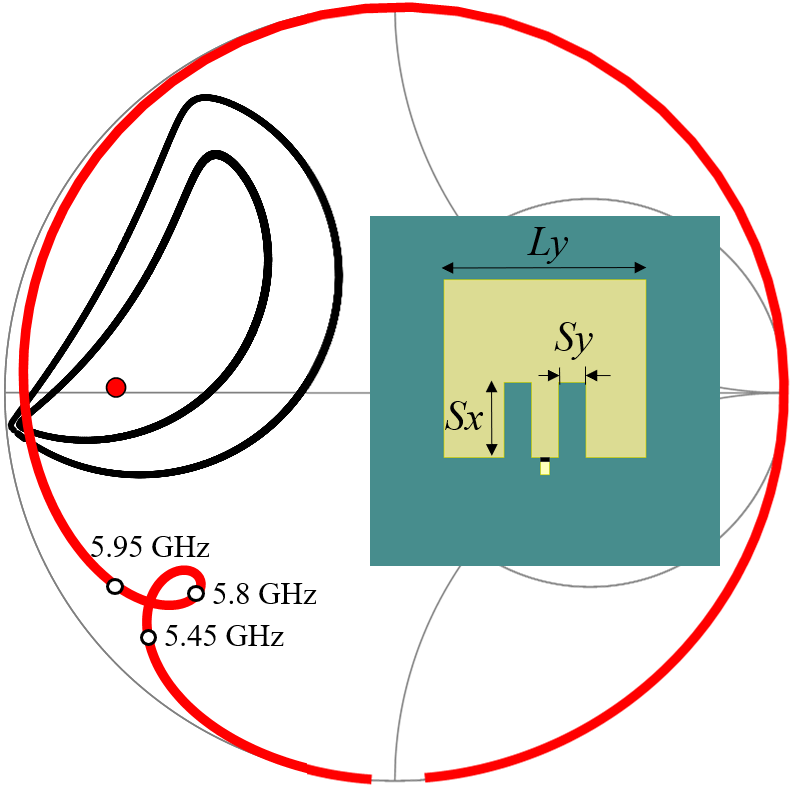} 
	} 
	\subfigure[] { 
		\includegraphics[width=0.45\columnwidth]{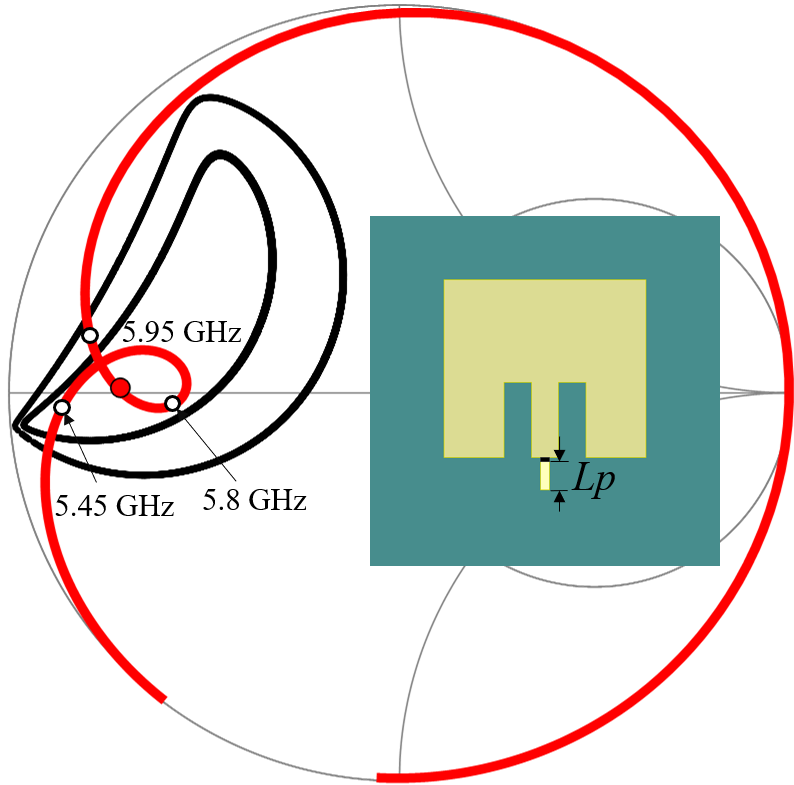} 
	} 
	\caption{The parameter tuning process of the C-band RRA element. (a) Step 0: The initial patch antenna. $Lx=11$ mm, $Ly=12$ mm, $Lp=1$ mm. (b) Step 1: Adjust the length of the patch $Lx$ to $Lx=13$ mm. (c) Step 2: Etch slots with $Sx=5.5$ mm, $Sy=2$ mm, and adjust the width of the patch $Ly$ to $Ly=14.8$ mm. (d) Step 3: Adjust the length of the pad $Lp$ to $Lp=2.1$ mm.}
	\label{paratune}
\end{figure}

During the element design, the parameter tuning process of the structure follows the wideband impedance matching technique employed in classical radiation antennas. Notably, the target matching point is the $S^t_{22}$ in Fig. \ref{58Gpara}, instead of the conventional 50 $\Omega$ impedance. Fig. \ref{paratune} shows the step-by-step parameter tuning process. Initially, we begin with a classical patch antenna, as depicted in Fig. \ref{paratune}(a). It is found that the resonant loop is higher than the desired frequency 5.8 GHz. Consequently, the patch length is extended so that 5.8 GHz locates near the center of the resonant loop, as shown in Fig. \ref{paratune}(b). However, the loop size is relatively large, thus failing to meet the bandwidth requirement. To reduce the loop size and thus increase the bandwidth, slots are etched to form a classical E-shaped structure and the width of the patch is also increased, as illustrated in Fig. \ref{paratune}(c). To further shift the curve towards the matching point, the length of the pad is increased to improve the serial inductance, as shown in Fig. \ref{paratune}(d). Finally, through these sequential parameter tuning steps, the curve is matched to the design target, resulting in a wide bandwidth.

\subsection{Results}
\label{result}
\begin{figure}[!t]
	\centerline{\includegraphics[width=0.6\columnwidth]{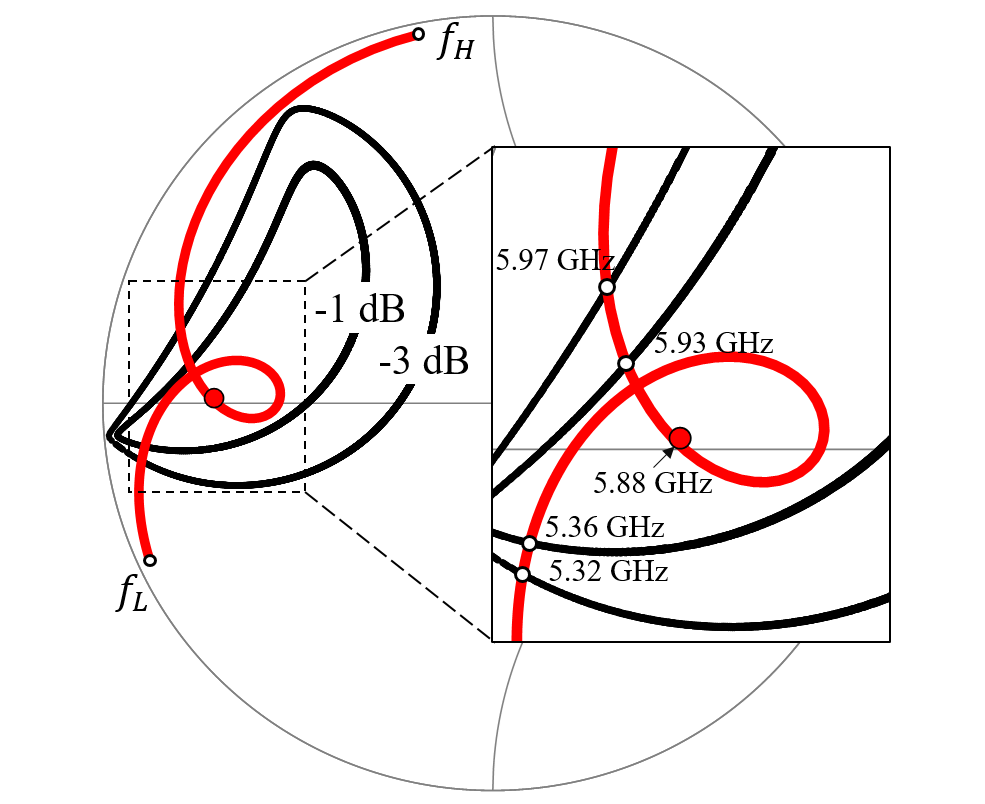}}
	\caption{The simulated $S_{22}$ curve of the structure (red curve) and the pre-calculated parameters in the Smith chart, where $f_L=$ 5 GHz and $f_H=$ 6.2 GHz.}
	\label{58GS22}
\end{figure}

With the lumped port at the gap to excite the patch, the reflection coefficient from the lumped port ($S_{22}$) is obtained from 5 GHz to 6.2 GHz, as plotted in the Smith chart (Fig. \ref{58GS22}). The curve is designed to vary near the design target over a wide bandwidth, and it passes through the design target at 5.88 GHz, indicating that a high-performance element can be obtained at this frequency. Moreover, the curve varies within the 1-dB CLC from 5.36 GHz to 5.93 GHz, and in other words, the 1-dB band is from 5.36 GHz to 5.93 GHz. Similarly, the 3-dB operation band is from 5.32 GHz to 5.97 GHz.

\subsection{Verification}

\begin{figure} 
	\centering 
	\subfigure[] { \label{58Gamp} 
		\includegraphics[width=0.45\columnwidth]{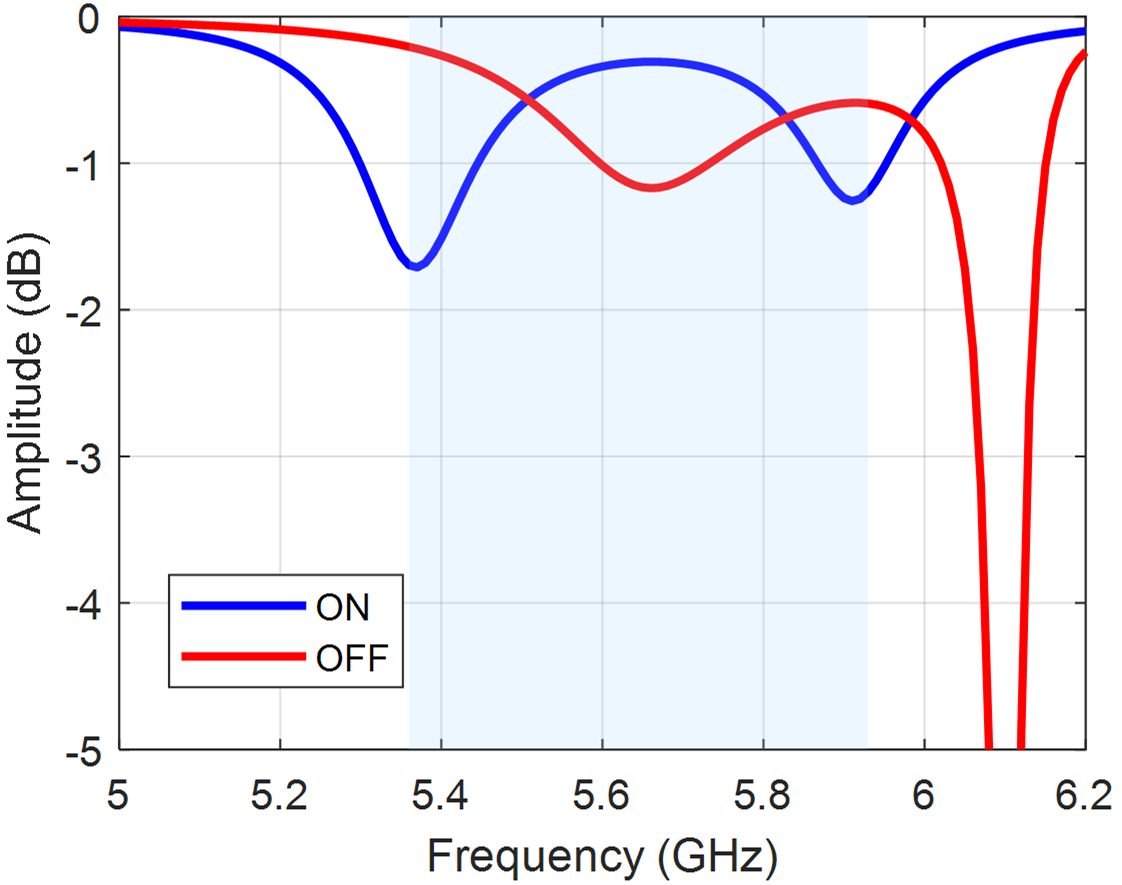} 
	} 
	\subfigure[] { \label{58Gphase} 
		\includegraphics[width=0.45\columnwidth]{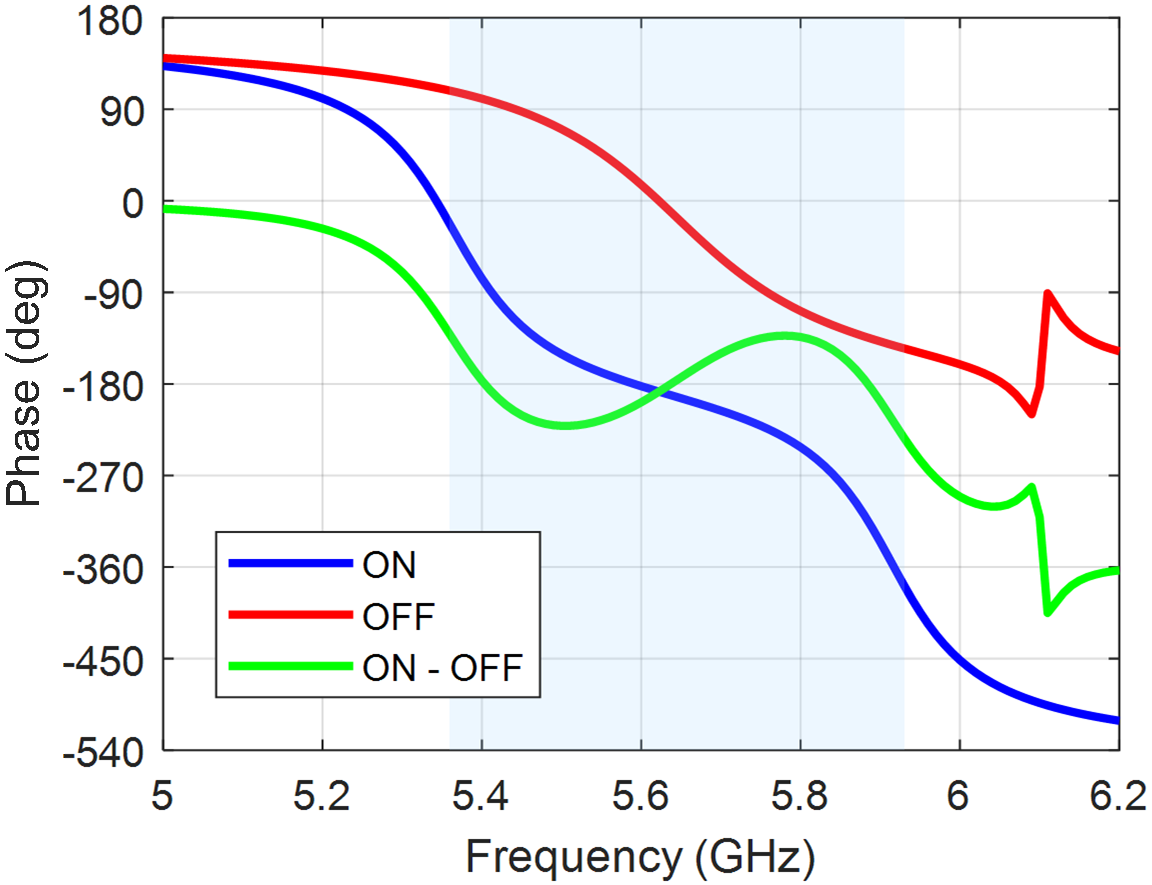} 
	} 
	\caption{The simulation verification from the conventional scattering method. (a) Reflection amplitudes. (b) Reflection phases under two switch states.} 
	\label{58Gper} 
\end{figure}

The conventional scattering method is used to verify this design. The lumped port is replaced with the RLC series model of the PIN diode, and the element is straightforwardly simulated with the Floquet port as the excitation source. As shown in Fig. \ref{58Gper}, the proposed E-shaped element is indeed a wideband 1-bit RRA element design, as there are multiple resonances over the desired band. The reflection amplitudes are higher than -1.8 dB from 5.36 GHz to 5.93 GHz, and the minimum amplitude loss at the cross point of two reflection amplitude curves is 0.58 dB, which is close to the calculated PL of 0.3 dB. The reflection phase varies within 180$^\circ\pm$50$^\circ$ over the band.

\begin{figure}[!t]
	\centerline{\includegraphics[width=0.6\columnwidth]{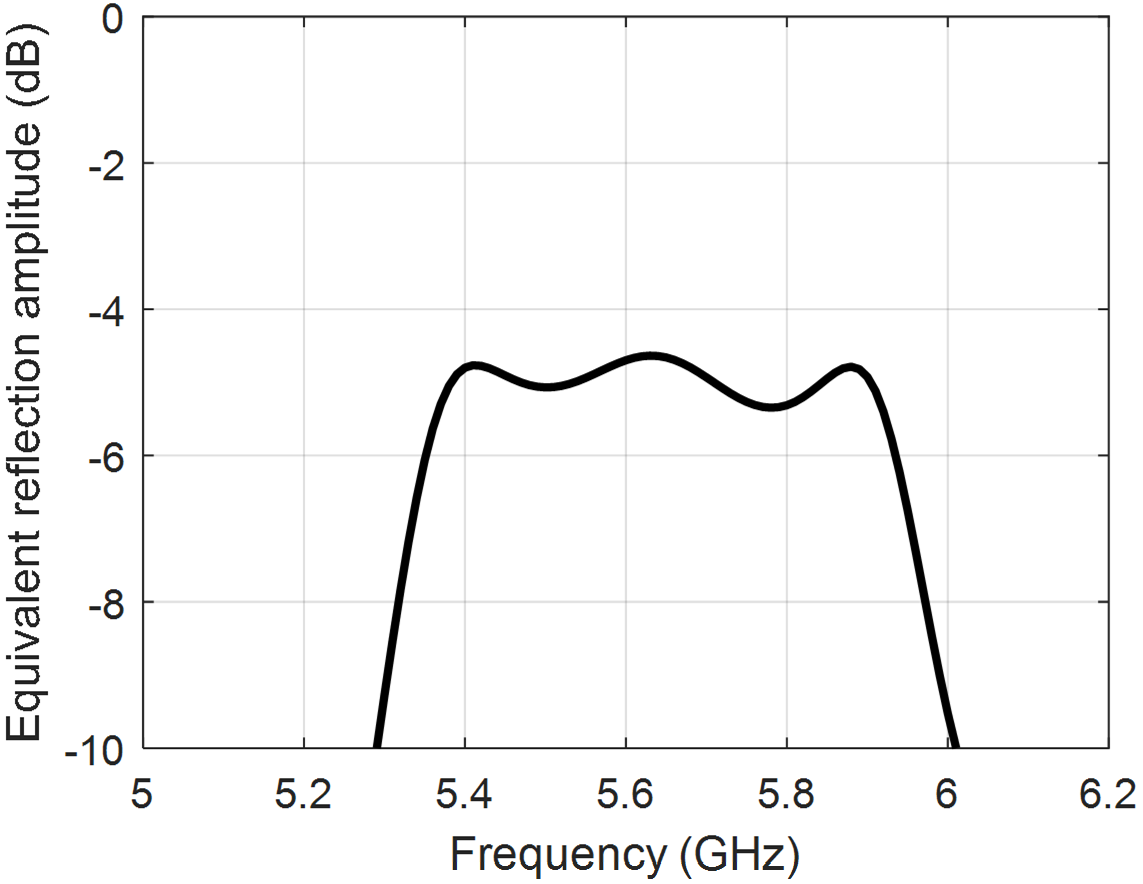}}
	\caption{The post-calculated ERA based on the simulated reflection coefficients from the Floquet port.}
	\label{58Gpost}
\end{figure}

Directly inserting the simulated reflection coefficients $\Gamma_\text{ON}$ and $\Gamma_\text{OFF}$ into (\ref{eq12}), the ERA can be post-calculated for quantitative performance evaluation. As shown in Fig. \ref{58Gpost}, the ERA reaches its maximum of -4.63 dB at 5.63 GHz, and at 5.88 GHz the post-calculated ERA is -4.79 dB, while the pre-calculated PL is -4.2 dB. The simulated maximum ERA is near and within the PL, with the discrepancies due to substrate loss and numerical simulation errors. The post-calculated 1-dB ERA bandwidth is from 5.36 GHz to 5.93 GHz (10.0\% fractional bandwidth), and the 3-dB ERA bandwidth is from 5.32 GHz to 5.97 GHz (11.5\% fractional bandwidth). Furthermore, the post-calculated ERA bandwidth matches the designed bandwidth in Section \ref{result}, demonstrating the effectiveness of the design using the proposed radiation viewpoint.

\section{Conclusion}

This article proposes a novel radiation viewpoint to model, evaluate and design the RRA element, aiming to address the challenges when designing wideband, multiband, or high-frequency RRA elements using the conventional scattering viewpoint. By decoupling the RRA element into a passive antenna structure and an active lumped switch, the two-port network of the antenna structure is used for analysis. To evaluate the element performance, ERA is defined, and the element PL, design target, and the CLCs can be pre-generated once the switch parameters are given, which can provide guidance when designing antennas. The proposed method is validated by a practical design example at THz band. Besides, the novel design process is applied to design a wideband 1-bit RRA element.

The proposed method can predict the element performance at given switch parameters before designing a specific element, which helps designers to evaluate and select proper switches to meet the performance requirements before carrying out actual element design. Moreover, using the radiation point of view, the art of designing RRA elements is the same as the mature impedance matching art of classical radiation antennas, which offers a systematic design method and significantly saves time during the conventional trial-and-error simulation process. In particular, wideband and multiband RRA elements can be designed intuitively with a clear and unique design target, and THz RRA elements can be designed and optimized to meet the PL using the imperfect switches at high frequency. 

Furthermore, the idea of element decoupling, network equivalence and radiation viewpoint can be extended to other active metasurfaces besides single-switch 1-bit RRA elements, such as 2-bit RRA elements, analog RRAs, lossy-substrate RRAs, energy-amplifying reflectarray antennas, dual-polarized RRAs, RTAs, and so on.

\appendices

\section{Derivations of the Equivalent Reflection Amplitude (ERA) Definition}
\label{appdA}

To maximize the gain of the array statistically, the element objective function should consider both the reflection amplitudes and quantized phases. Previous research in \cite{bit2} found that the incident phase on each RRA element is pseudorandom. To derive an expression of ERA theoretically, it is further supposed that the each element has a random incident phase $\phi$ with a uniform incident amplitude. Additionally, suppose the size of array is infinite with an element number of $M\to \infty$. The total reflection coefficient is the summation of each element
\begin{equation}
	E=\sum_{m=1}^M A_m e^{-j\left(\varphi_m-\phi_m\right)},
\end{equation}
where $\varphi_m$ is the additional phase of the $m$th element, and $A_m$ is the $m$th reflection amplitude. To maximize the gain, the $m$th reflection coefficient is selected to match with the incident phase $\phi_m$ and form in-phase addition.

\begin{figure}[!t]
	\centerline{\includegraphics[width=0.5\columnwidth]{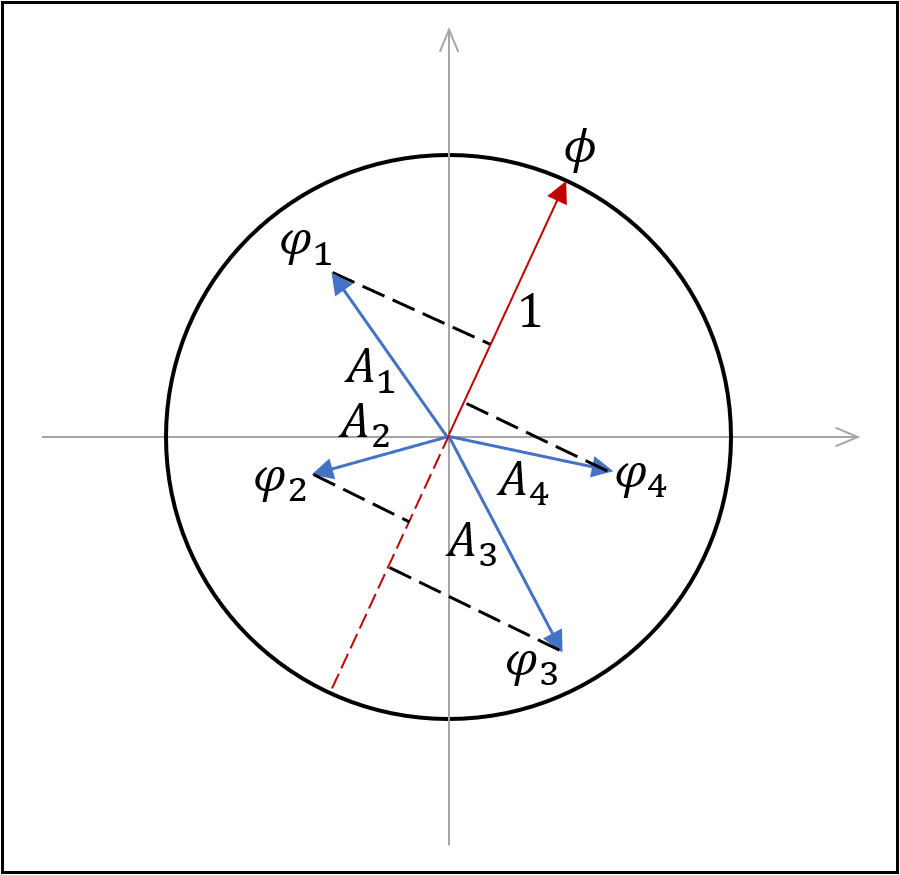}}
	\caption{A conceptual illustration of state projection and quantization strategy of an element. The blue arrows denote the element states, and the red arrow denotes the random incident vector. All the element states are projected onto the incident vector. In this 4-state case, the state 1 has the maximum projection value, so the incident vector should be quantized to state 1.}
	\label{Aeqproof}
\end{figure}

Suppose each element has $N$ given reflection states, namely $\Gamma_i = A_ie^{j\varphi_i}$, where $i\in {1, ... , N}$, and then a quantization strategy is adopted to maximize the array gain. In Fig. \ref{Aeqproof}, we show how to quantize the states of a single element. First, all the state vectors are projected onto the incident vector in the polar plot, where the incident vector has a random phase $\phi$ and uniform amplitude. Then the incident vector is quantized to the state with the maximum projection value. Mathematically, the quantization strategy is expressed as follows
\begin{equation}
	\mathrm{State} = \underset{i}{\operatorname{argmax}} \{A_i\cos(\phi-\varphi_i)\}\quad i\in {1, ... , N},
\end{equation}
where $\cos$ means projecting the incident vector to the given states. On the other hand, the maximum projection value
\begin{equation}
	P_\phi=\max \limits_{i}\{A_i\cos(\phi-\varphi_i)\}\quad i\in {1, ... , N}
\end{equation}
presents the ERA of a single element after quantization. Since the incident phase is a random variable with uniform distribution in $[0, 2\pi]$, we obtain the total ERA of all elements by taking the average of their separate ERAs. The average ERA is expressed as follows 
\begin{equation}
	\text{ERA}=\frac{1}{2\pi}\int_{0}^{2\pi}\max \limits_{i}\{A_i\cos(\phi-\varphi_i)\} \mathrm{d} \phi.
\end{equation}
ERA includes both reflection amplitude and phase. In an ideal case where the reflection amplitudes of all states are 1, and the reflection phase can be chosen continuously, the ERA is 1, which means there is no element reflection loss.

For ideal 1-bit phase quantization case, the reflection amplitudes at ON/OFF states are 1, and the reflection phase difference is $180^\circ$. The ERA can be calculated as follows
\begin{equation}
	\text{ERA}=\frac{1}{2\pi}\cdot 2\int_{-\frac{\pi}{2}}^{\frac{\pi}{2}}\cos(\phi) \mathrm{d} \phi = \frac{2}{\pi},
\end{equation}
which is equivalent to -3.9 dB. This equivalent loss is solely due to the 1-bit phase quantization effect. Besides, it is noted that this quantization loss of around 3.9 dB has been observed in previous literature \cite{bit1}, which supports the validity of the proposed quantization strategy and ERA definition.

\section{Analytical Solution of the Design Target under 1-Bit Case}
\label{appdB}

With the observation of the optimal reflection coefficients under 1-bit RRA case, it is found that the simplified design condition is $\Gamma_\text{ON}=-\Gamma_\text{OFF}$. Using this condition, an analytical solution for $S_{22}^t$ can be derived by solving the following equation
\begin{equation}
	\frac{|S_{22}|-e^{j\theta_{22}}\gamma_\text{ON}}{1-|S_{22}|e^{j\theta_{22}}\gamma_\text{ON}} = - \frac{|S_{22}|-e^{j\theta_{22}}\gamma_\text{OFF}}{1-|S_{22}|e^{j\theta_{22}}\gamma_\text{OFF}}.
	\label{eqana}
\end{equation}
(\ref{eqana}) is simplified into a binary quadratic equation
\begin{equation}
	|S_{22}|^2-2\frac{x_\text{ON}x_\text{OFF}+1}{x_\text{ON}+x_\text{OFF}}|S_{22}|+1=0,
	\label{eqbqe}
\end{equation}
where $x_\text{ON}=e^{j\theta_{22}}\gamma_\text{ON}$ and $x_\text{OFF}=e^{j\theta_{22}}\gamma_\text{OFF}$. Since $|S_{22}|$ is a real number, the conjugate form of (\ref{eqbqe}) is also valid
\begin{equation}
	|S_{22}|^2-2\frac{\overline{x_\text{ON}}\overline{x_\text{OFF}}+1}{\overline{x_\text{ON}}+\overline{x_\text{OFF}}}|S_{22}|+1=0.
	\label{eqbqec}
\end{equation}
By subtracting (\ref{eqbqe}) and (\ref{eqbqec}), $|S_{22}|^2$ term is eliminated
\begin{equation}
	\left(\frac{x_\text{ON}x_\text{OFF}+1}{x_\text{ON}+x_\text{OFF}}-\frac{\overline{x_\text{ON}}\overline{x_\text{OFF}}+1}{\overline{x_\text{ON}}+\overline{x_\text{OFF}}}\right)|S_{22}|=0,
\end{equation}
where $|S_{22}|\ne 0$ in general case. So the following equation is obtained without $|S_{22}|$
\begin{equation}
	\frac{x_\text{ON}x_\text{OFF}+1}{x_\text{ON}+x_\text{OFF}}=\frac{\overline{x_\text{ON}}\overline{x_\text{OFF}}+1}{\overline{x_\text{ON}}+\overline{x_\text{OFF}}},
\end{equation}
and rearranged as
\begin{equation}
	\begin{aligned}
		&\left(1-|x_\text{OFF}|^2\right) x_\text{ON}+\left(1-|x_\text{ON}|^2\right) x_\text{OFF}\\
		=&\left(1-|x_\text{OFF}|^2\right) \overline{x_\text{ON}}+\left(1-|x_\text{ON}|^2\right) \overline{x_\text{OFF}}.
		\label{eqtheta}
	\end{aligned}
\end{equation}
Inserting $x_\text{ON}=e^{j\theta_{22}}\gamma_\text{ON}$ and $x_\text{OFF}=e^{j\theta_{22}}\gamma_\text{OFF}$ into (\ref{eqtheta}), the solution of $e^{j \theta_{22}}$ is
\begin{equation}
	e^{j \theta_{22}} =\pm \sqrt{\frac{\left(1-|\gamma_\text{ON}|^2\right) \overline{\gamma_\text{OFF}}+\left(1-|\gamma_\text{OFF}|^2\right) \overline{\gamma_\text{ON}}}{\left(1-|\gamma_\text{ON}|^2\right) \gamma_\text{OFF}+\left(1-|\gamma_\text{OFF}|^2\right) \gamma_\text{ON}}}.
	\label{eqsolvtheta22}
\end{equation}
Solving (\ref{eqbqe}), we obtain
\begin{equation}
	|S_{22}|=\frac{x_\text{ON}x_\text{OFF}+1-\sqrt{\left(1-x_\text{OFF}^2\right)\left(1-x_\text{ON}^2\right)}}{x_\text{ON}+x_\text{OFF}}.
	\label{eqs22}
\end{equation}
Inserting $e^{j \theta_{22}}$ in (\ref{eqsolvtheta22}) into (\ref{eqs22}), the solution of $|S_{22}|$ is
\begin{strip}
	\begin{equation}
		|S_{22}| = \frac{\left|1-|\gamma_\text{ON}|^2|\gamma_\text{OFF}|^2-\sqrt{\left(1-|\gamma_\text{ON}|^2\right)\left(1-|\gamma_\text{OFF}|^2\right)\left(\gamma_\text{ON} \overline{\gamma_\text{OFF}}-1\right)\left(\gamma_\text{OFF} \overline{\gamma_\text{ON}}-1\right)}\right|}{\left|\left(1-|\gamma_\text{ON}|^2\right) \gamma_\text{OFF}+\left(1-|\gamma_\text{OFF}|^2\right) \gamma_\text{ON}\right|}.
		\label{eqsolvs22}
	\end{equation}
\end{strip}

(\ref{eqsolvtheta22}) and (\ref{eqsolvs22}) are the analytical solutions for $S_{22}^t$, which match with the values of $S_{22}^t$ obtained using the traversal numerical method.

\begin{IEEEbiography}[{\includegraphics[width=1in,height=1.25in,clip,keepaspectratio]{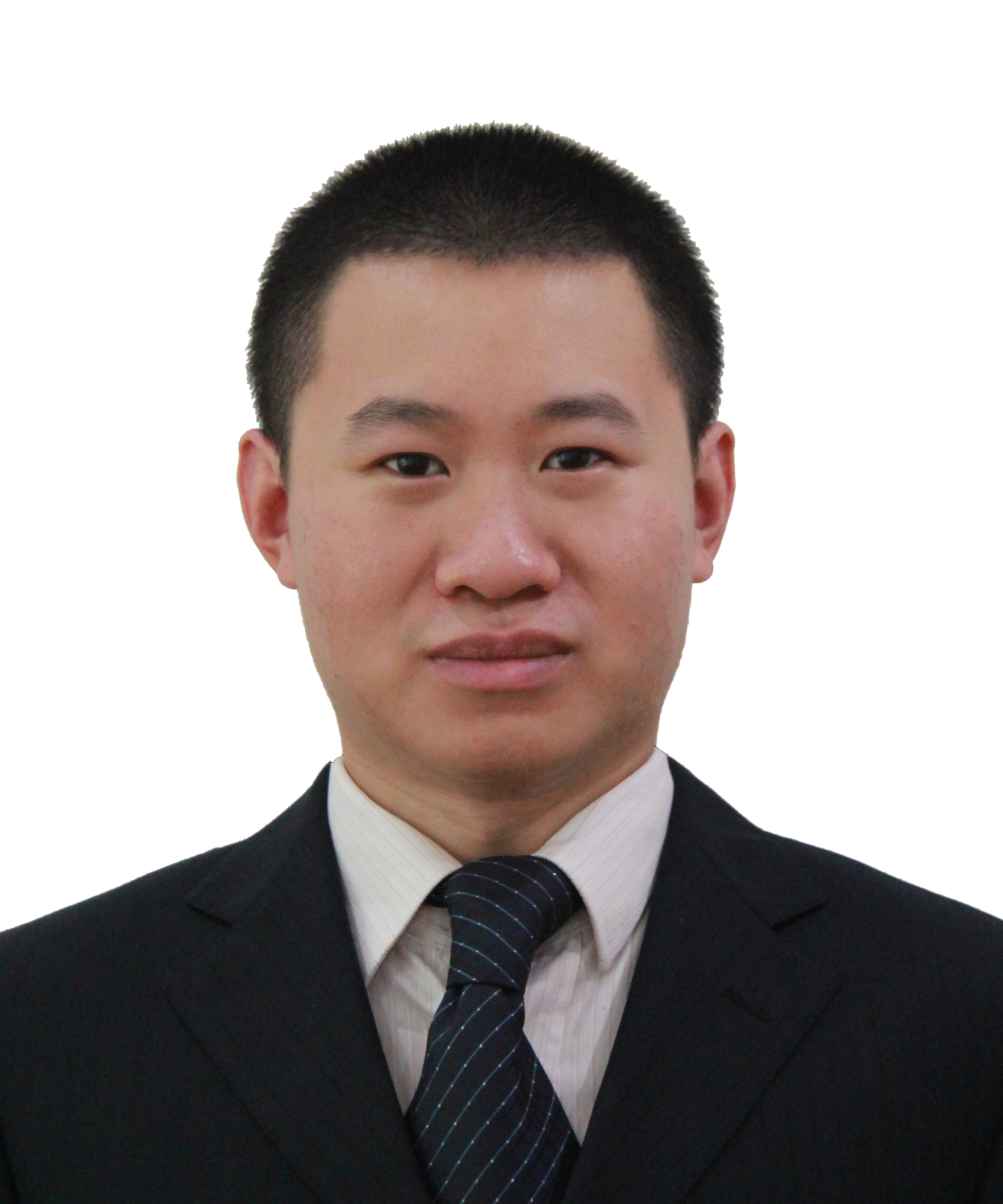}}]{Changhao Liu} (Student Member, IEEE)
	received the B.S. degree in electronic engineering from Tsinghua University, Beijing, China, in 2021, where he is currently pursuing the Ph.D. degree in electronic engineering.
	
	From 2019 to 2021, he was a Research Assistant at the Microwave and Antenna Institute, Department of Electronic Engineering, Tsinghua University. His current research interests include reconfigurable metasurfaces, surface electromagnetics, reflectarray antennas, transmitarray antennas, and terahertz metasurfaces.
\end{IEEEbiography}

\begin{IEEEbiography}[{\includegraphics[width=1in,height=1.25in,clip,keepaspectratio]{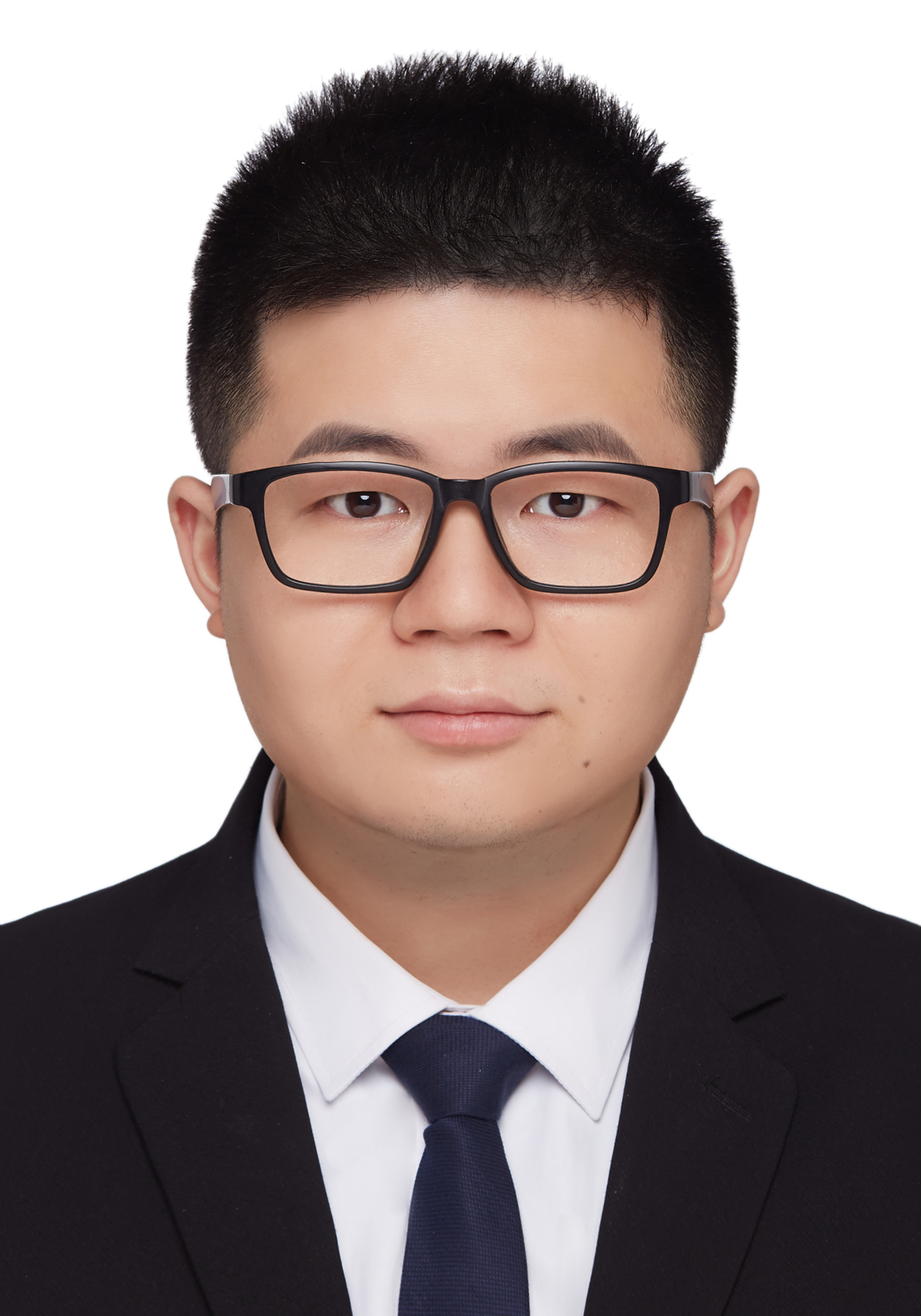}}]{You Wu}
	received the B.S. degree and the M.S. degree from Beihang University, Beijing, China, in 2016 and in 2019, respectively. He is currently pursuing the Ph.D.degree with the Department of Electronic Engineering, Tsinghua University, Beijing. 
	
	His current research interests include reconfigurable reflectarray, W band antenna, and terahertz antenna.
\end{IEEEbiography}

\begin{IEEEbiography}[{\includegraphics[width=1in,height=1.25in,clip,keepaspectratio]{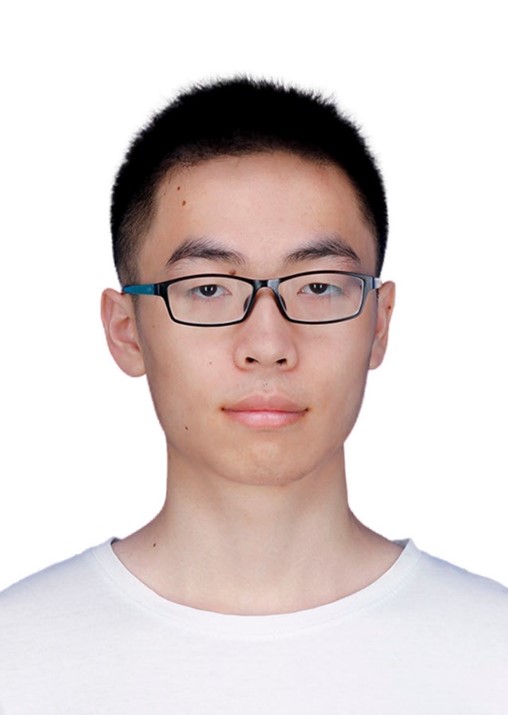}}]{Songlin Zhou}
	received the B.S. degree from Tsinghua University, Beijing, China, in 2020. He is currently pursuing the Ph.D. degree with the Department of Electronic Engineering, Tsinghua University, Beijing.
	
	His current research interests include reconfigurable reflectarray, multi-polarized reflectarray and angular multiplexing in reflectarray.

\end{IEEEbiography}

\begin{IEEEbiography}[{\includegraphics[width=1in,height=1.25in,clip,keepaspectratio]{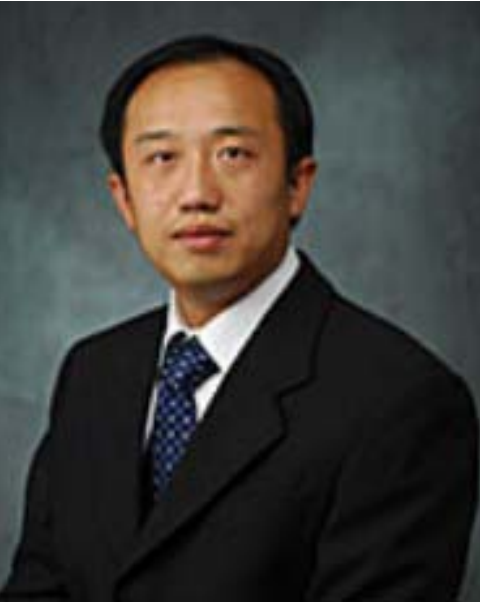}}]{Fan Yang} (Fellow, IEEE) received the B.S. and M.S.
	degrees from Tsinghua University, Beijing, China,
	in 1997 and 1999, respectively, and the Ph.D. degree
	from the University of California at Los Angeles
	(UCLA), Los Angeles, CA, USA, in 2002.
	
	From 1994 to 1999, he was a Research Assistant
	at the State Key Laboratory of Microwave and
	Digital Communications, Tsinghua University. From
	1999 to 2002, he was a Graduate Student Researcher
	at the Antenna Laboratory, UCLA. From 2002 to
	2004, he was a Post-Doctoral Research Engineer and
	an Instructor at the Department of Electrical Engineering, UCLA. In 2004,
	he joined the Department of Electrical Engineering, The University of
	Mississippi, Oxford, MS, USA, as an Assistant Professor, where he was
	promoted to an Associate Professor in 2009. In 2011, he joined the Department
	of Electronic Engineering, Tsinghua University, as a Professor, where he has
	been the Director of the Microwave and Antenna Institute since 2011. He has
	published over 300 journal articles and conference papers, 6 book chapters,
	and 5 books entitled \textit{Reflectarray Antennas: Theory, Designs, and Applications} (IEEE-Wiley, 2018), \textit{Analysis and Design of Transmitarray Antennas}
	(Morgan \& Claypool, 2017), \textit{Scattering Analysis of Periodic Structures
		Using Finite-Difference Time-Domain Method} (Morgan \& Claypool, 2012),
	\textit{Electromagnetic Band Gap Structures in Antenna Engineering} (Cambridge
	University Press, 2009), and \textit{Electromagnetics and Antenna Optimization
		Using Taguchi’s Method} (Morgan \& Claypool, 2007). His research interests
	include antennas, surface electromagnetics, computational electromagnetics,
	and applied electromagnetic systems.
	
	Dr. Yang is a Fellow of the Applied Computational Electromagnetics
	Society (ACES). He was a recipient of several prestigious awards and
	recognitions, including the Young Scientist Award of the 2005 URSI General
	Assembly and the 2007 International Symposium on Electromagnetic Theory,
	the 2008 Junior Faculty Research Award of the University of Mississippi,
	the 2009 inaugural IEEE Donald G. Dudley Jr. Undergraduate Teaching
	Award, and the 2011 Recipient of Global Experts Program of China. He was
	the Technical Program Committee (TPC) Chair of the 2014 IEEE International
	Symposium on Antennas and Propagation and the USNC-URSI Radio Science
	Meeting. He served as an Associate Editor for the {\scshape IEEE Transactions on
		Antennas and Propagation} from 2010 to 2013 and the Associate Editor-in-Chief for \textit{Applied Computational Electromagnetics Society (ACES) Journal}
	from 2008 to 2014. He is a Distinguished Lecturer of IEEE Antennas and
	Propagation Society (APS), since 2018.\end{IEEEbiography}

	\begin{IEEEbiography}[{\includegraphics[width=1in,height=1.25in,clip,keepaspectratio]{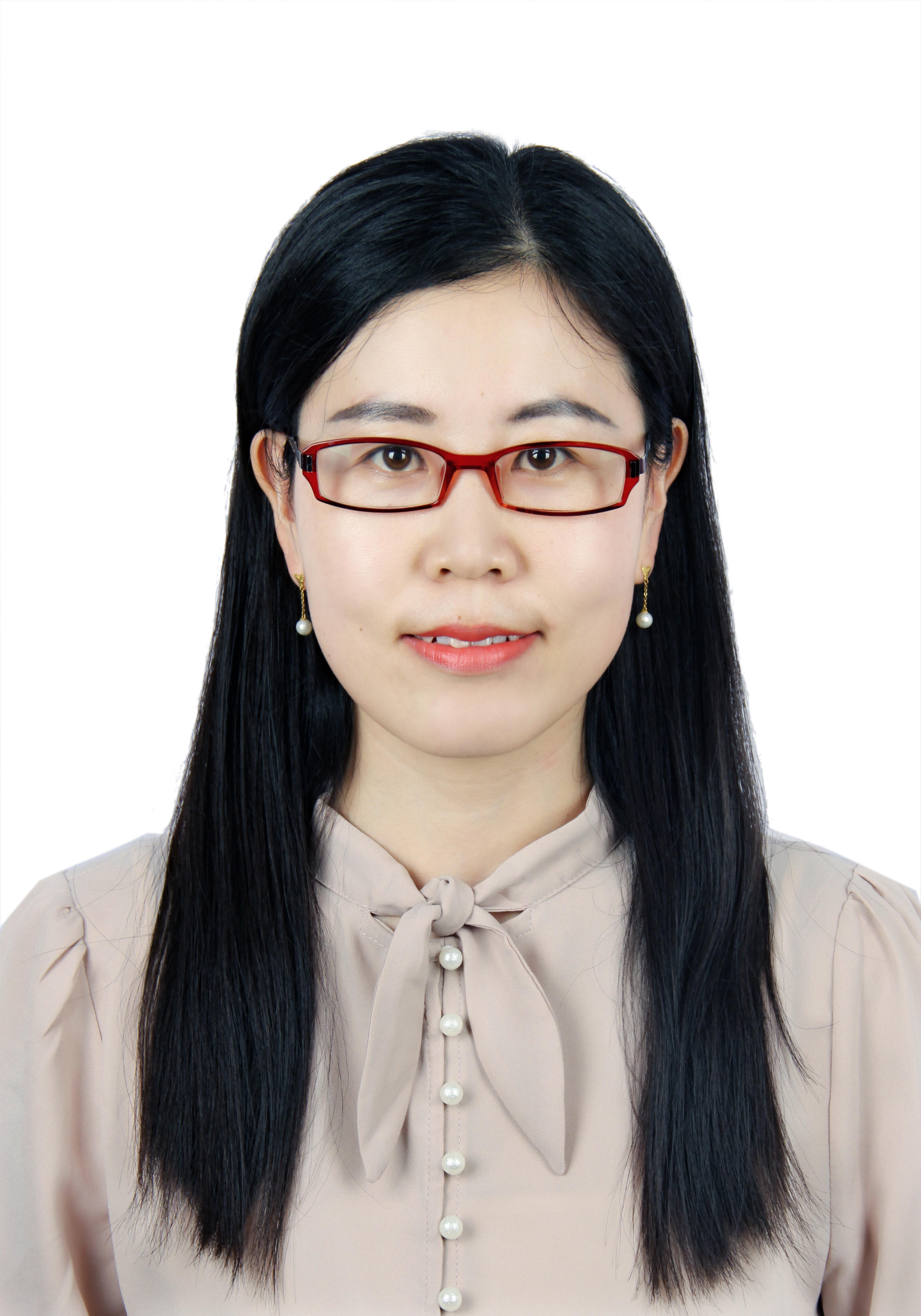}}]{Rongli Ren}
		received the B.S. degree from Chengdu University of Information Technology, Chengdu, China, in 2010, and the M.S. degree from the University of Electronic Science and Technology of China, Chengdu, China, in 2014. 
		
		She is currently an engineer at the Department of Electronic Engineering, Tsinghua University. Her main research interests include theoretical research and design of millimeter-wave and terahertz high-performance intelligent metasurfaces, as well as wireless communication system antenna measurement methods and transmission system verification.
	\end{IEEEbiography}

\begin{IEEEbiography}[{\includegraphics[width=1in,height=1.25in,clip,keepaspectratio]{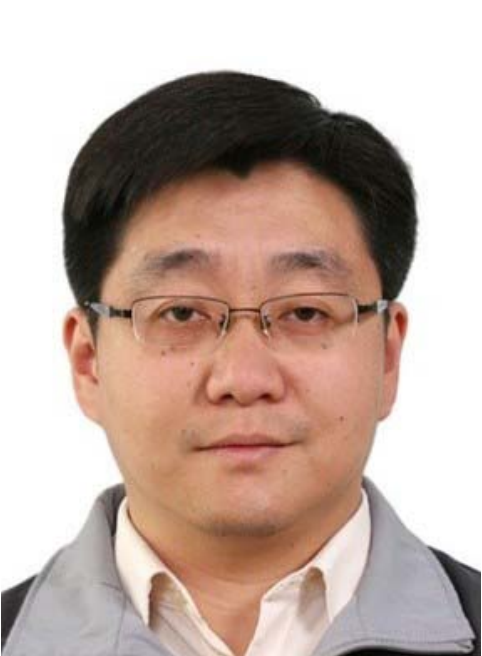}}]{Shenheng Xu} (Member, IEEE)
	received the B.S.
	and M.S. degrees from the Southeast University,
	Nanjing, China, in 2001 and 2004, respectively, and
	the Ph.D. degree in electrical engineering from the
	University of California at Los Angeles (UCLA),
	Los Angeles, CA, USA, in 2009.
	
	From 2000 to 2004, he was a Research Assistant
	with the State Key Laboratory of Millimeter Waves,
	Southeast University. From 2004 to 2011, he was
	a Graduate Student Researcher and later a Post-Doctoral Researcher with the Antenna Research,
	Analysis, and Measurement Laboratory, UCLA. In 2012, he joined the
	Department of Electronic Engineering, Tsinghua University, Beijing, China,
	as an Associate Professor. His research interests include novel designs of high-gain antennas for advanced applications, artificial electromagnetic structures,
	and electromagnetic and antenna theories.
\end{IEEEbiography}

\begin{IEEEbiography}[{\includegraphics[width=1in,height=1.25in,clip,keepaspectratio]{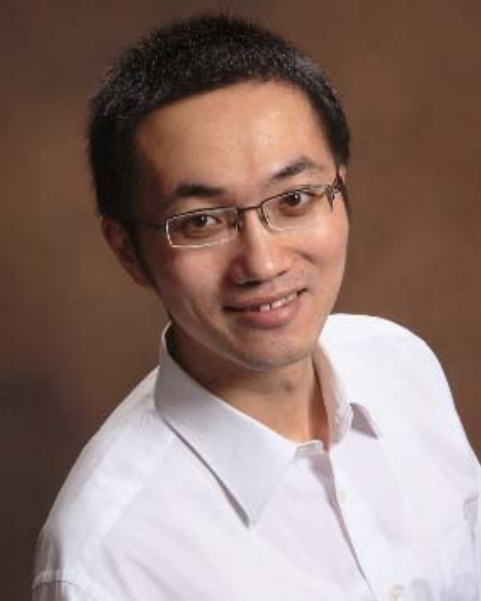}}]{Maokun Li} (Senior Member, IEEE)
	received the
	B.S. degree in electronic engineering from Tsinghua
	University, Beijing, China, in 2002, and the M.S.
	and Ph.D. degrees in electrical engineering from the
	University of Illinois at Urbana–Champaign, Champaign, IL, USA, in 2004 and 2007, respectively.
	
	After graduation, he worked as a Senior Research
	Scientist with Schlumberger-Doll Research, Cambridge, MA, USA. Since 2014, he has been with
	the Department of Electronic Engineering, Tsinghua
	University. He has authored or coauthored one book
	chapter, 50 journal articles, and 120 conference proceedings. He holds
	three patent applications. His research interests include fast algorithms in
	computational electromagnetics and their applications in antenna modeling,
	electromagnetic compatibility analysis, inverse problems, and so on.
	
	Dr. Li was a recipient of the China National 1000 Plan in 2014 and the
	2017 IEEE Ulrich L. Rohde Innovative Conference Paper Award. He serves
	as an Associate Editor for the {\scshape IEEE Journal on Multiscale and Multiphysics Computational Techniques} and \textit{Applied Computational
		Electromagnetic Society Journal} and a Guest Editor for the Special Issue
	on Electromagnetic Inverse Problems for Sensing and Imaging of the \textit{IEEE
		Antennas and Propagation Magazine}.
\end{IEEEbiography}

\end{document}